\documentclass[11pt,peereview,draftclsnofoot,onecolumn]{IEEEtran}

\usepackage{amsmath,amssymb,graphicx,psfrag}

\def\bfy{{\mathbf{y}}}
\def\bfH{{\mathbf{H}}}
\def\bfh{{\mathbf{h}}}
\def\bfs{{\mathbf{s}}}
\def\bfsh{{\hat{\mathbf{s}}}}

\def\bfv{{\mathbf{v}}}
\def\bfx{{\mathbf{x}}}
\def\bfX{{\mathbf{X}}}
\def\bfL{{\mathbf{L}}}
\def\bfe{{\mathbf{e}}}
\def\bfZero{{\mathbf{0}}}
\def\bfM{{\mathbf{M}}}
\def\bfI{{\mathbf{I}}}
\def\bfQ{{\mathbf{Q}}}
\def\bfY{{\mathbf{Y}}}
\def\bfYt{\tilde{\mathbf{Y}}}
\def\bfYb{\bar{\mathbf{Y}}}
\def\bfA{{\mathbf{A}}}
\def\bfAt{\tilde{{\mathbf{A}}}}
\def\bfAh{\hat{{\mathbf{A}}}}
\def\bfAt{\tilde{{\mathbf{A}}}}
\def\bfa{{\mathbf{a}}}
\def\bfB{{\mathbf{B}}}
\def\bfb{{\mathbf{b}}}
\def\bfw{{\mathbf{w}}}
\def\bfu{{\mathbf{u}}}
\def\bfub{\bar{\mathbf{u}}}
\def\bfuh{\hat{\mathbf{u}}}
\def\bfut{\tilde{\mathbf{u}}}
\def\bfU{{\mathbf{U}}}
\def\bfUb{{\bar{\mathbf{U}}}}
\def\bfUh{{\hat{\mathbf{U}}}}
\def\bfUt{{\tilde{\mathbf{U}}}}
\def\bfSigma{{\mathbf{\Sigma}}}
\def\bfd{{\mathbf{d}}}
\def\bfP{{\mathbf{P}}}
\def\bfT{{\mathbf{T}}}
\def\bfR{{\mathbf{R}}}
\def\bflambda{{\boldsymbol{\lambda}}}
\def\bflambdab{\bar{{\boldsymbol{\lambda}}}}
\def\bflambdah{\hat{{\boldsymbol{\lambda}}}}
\def\bflambdat{\tilde{{\boldsymbol{\lambda}}}}
\def\bfLambda{{\mathbf{\Lambda}}}
\def\bfLambdab{\bar{\mathbf{\Lambda}}}
\def\bfLambdah{\hat{\mathbf{\Lambda}}}
\def\bfLambdat{\tilde{\mathbf{\Lambda}}}
\def\bfz{{\mathbf{z}}}
\def\bfzb{{\bar{\mathbf{z}}}}
\def\bfzh{{\hat{\mathbf{z}}}}
\def\bfzt{{\tilde{\mathbf{z}}}}
\def\bfomega{\boldsymbol{\omega}}
\def\bfomegab{\bar{\boldsymbol{\omega}}}
\def\bfxi{\boldsymbol{\xi}}
\def\bfxih{\hat{\boldsymbol{\xi}}}
\def\bfxit{\tilde{\boldsymbol{\xi}}}
\def\bfD{{\mathbf{D}}}
\def\bfDt{\tilde{\mathbf{D}}}
\def\bfgamma{\boldsymbol{\gamma}}
\def\bfgammah{\hat{\boldsymbol{\gamma}}}
\def\bfgammat{\tilde{\boldsymbol{\gamma}}}
\def\bfg{{\mathbf{g}}}
\def\bfgb{\bar{\mathbf{g}}}

\def\bfGb{\bar{\mathbf{G}}}
\def\bfF{{\mathbf{F}}}
\def\bfFb{{\bar{\mathbf{F}}}}
\def\bftheta{\boldsymbol{\theta}}

\def\bfthetat{\tilde{\boldsymbol{\theta}}}
\def\reals{{\mathbb{R}}}
\def\complex{{\mathbb{C}}}
\def\bins{{\mathcal{B}}}
\def\sym{{\mathbb{S}}}
\def\Xset{{\mathcal{X}}}
\def\Yset{{\mathcal{Y}}}
\def\Hset{{\mathcal{H}}}
\def\Ysett{\tilde{\mathcal{Y}}}
\def\Aset{{\mathcal{A}}}
\def\Mset{{\mathcal{M}}}
\def\Nset{{\mathcal{N}}}
\def\Iset{{\mathcal{I}}}
\def\Uset{{\mathcal{U}}}
\def\Dset{{\mathcal{D}}}
\def\Pset{{\mathcal{P}}}
\def\Rset{{\mathcal{R}}}
\def\Cset{{\mathcal{C}}}
\newcommand{\Trace}{\mathrm{Tr}}
\newcommand{\diag}{\mathrm{diag}}
\newcommand{\Diag}{\mathrm{Diag}}
\newcommand{\rank}{\mathrm{Rank}}
\newcommand{\prob}[1]{\mathrm{P}\left(#1\right)}
\newcommand{\expt}[2][]{\mathrm{E}_{#1}\left\{ #2 \right\}}
\def\tr{^{\mathrm T}}
\def\defeq{\triangleq}
\def\bigO{{O}}
\def\SNR{\rho}
\def\svec{\mathrm{svec}}
\newtheorem{theorem}{Theorem}
\newtheorem{lemma}{Lemma}
\DeclareMathOperator*{\dotleq}{\overset{.}{\leq}}
\DeclareMathOperator*{\dotgeq}{\overset{.}{\geq}}

\begin{document}
\sloppy

\title{The Diversity Order of the \\ Semidefinite Relaxation Detector}

\author{\authorblockN{Joakim Jald\'{e}n and Bj\"{o}rn Ottersten}\\
\authorblockA{Signal Processing Lab, School of Electrical Engineering,\\
KTH, Royal Institute of Technology,\\
Stockholm, Sweden \\
{\tt [joakim.jalden,bjorn.ottersten]@ee.kth.se}}
\thanks{Manuscript submitted June 13, 2006. The material in this paper will be presented in part at the Asilomar Conference on Signals, Systems, and Computers, Pacific Grove, CA, October 2006.}}

\markboth{SUBMITTED TO THE IEEE TRANSACTIONS ON INFORMATION THEORY}{}
\maketitle

\begin{abstract}
We consider the detection of binary (antipodal) signals transmitted in a spatially multiplexed fashion over a fading multiple-input multiple-output (MIMO) channel and where the detection is done by means of semidefinite relaxation (SDR). The SDR detector is an attractive alternative to maximum likelihood (ML) detection since the complexity is polynomial rather than exponential. Assuming that the channel matrix is drawn with i.i.d. real valued Gaussian entries, we study the receiver diversity and prove that the SDR detector achieves the maximum possible diversity. Thus, the error probability of the receiver tends to zero at the same rate as the optimal maximum likelihood (ML) receiver in the high signal to noise ratio (SNR) limit. This significantly strengthens previous performance guarantees available for the semidefinite relaxation detector. Additionally, it proves that full diversity detection is in certain scenarios also possible when using a non-combinatorial receiver structure.
\end{abstract}

\begin{keywords}
Semidefinite relaxation, diversity, MIMO, detection.
\end{keywords}

\section{Introduction}

Herein, we consider the detection of binary symbols transmitted over an $n$ by $m$ multiple-input multiple-output (MIMO) channel modelled according to
\begin{equation} \label{eq:data-model}
\bfy = \bfH\bfs+\bfv
\end{equation}
where $\bfs \in \bins^m \defeq \{\pm 1\}^m$, $\bfH \in \reals^{n \times m}$ and $\bfv,\bfy \in \reals^n$. In what follows, $\bfy$ is referred to as the vector of \emph{received signals}; $\bfH$ as the \emph{channel matrix}; $\bfs$ as the \emph{transmitted message}; and $\bfv$ as the additive \emph{noise} based on their physical interpretations in the digital communications context. The additive noise is assumed to be white and Gaussian with a variance of $\rho^{-1}$ per component. It will also be assumed that the channel matrix, $\bfH$, is known to the receiver and that all possible transmitted messages, $\bfs$, are equally likely.

The problem of detecting a vector of symbols (not necessarily binary) transmitted over a MIMO channel is of general interest as it arises frequently in digital communications. Examples include, but are not limited to, the multiuser detection problem in CDMA~\cite{Ver:98} and communications over a multiple antenna channel~\cite{TW:05}. However, while the detection problem is the same for many areas, the structure and assumptions regarding the channel matrix, $\bfH$, will typically differ depending on the specific context. In the interest of simplicity, we will assume that the channel matrix may be modelled using i.i.d.~Gaussian entries with zero mean and finite variance, an assumption motivated by the problem of wireless communication over a richly scattered fading multiple antenna channel~\cite{TW:05}. The signal to noise ratio (SNR) of the channel is equal to $\rho$ and we will focus on an analysis of the high SNR regime.

The maximum likelihood (ML) estimate of $\bfs$, $\bfsh_\text{ML}$, is well known to be given by
\begin{equation} \label{eq:ml-detector}
\bfsh_\mathrm{ML} = \arg \min_{\bfsh \in \bins^m} \|\bfy - \bfH\bfsh\|^2
\end{equation}
where $\|\cdot\|$ denotes the Euclidian norm, i.e. the ML detector, or receiver, selects the message, $\bfsh$, which minimizes the distance between the received signals and the hypothesized noise-free message, $\bfH\bfsh$. An error is declared whenever $\bfsh_\mathrm{ML} \neq \bfs$ and it well known that the ML detector is optimal in the sense that it minimizes the probability of error. However, for a general channel matrix, $\bfH$, and vector of received signals, $\bfy$, the ML detection problem in~\eqref{eq:ml-detector} has been shown to be NP-hard~\cite{Ver:89} and the full search solution has a complexity of $\bigO(2^m)$ where $m$ is the number of symbols jointly detected. A similar result holds for the sphere decoding algorithm which is able to provide exact solutions to~\eqref{eq:ml-detector} at an expected complexity on the order of $\bigO(2^{\gamma m})$ for some $\gamma \in (0,1]$~\cite{JO:04}. The complexity is thus, although significantly lower than the full search, still exponential.

\begin{figure}
\psfrag{b}[cc]{$\prob{\bfsh \neq \bfs}$}
\psfrag{m}[cc]{SNR [dB]}
\includegraphics[width=1.0\linewidth]{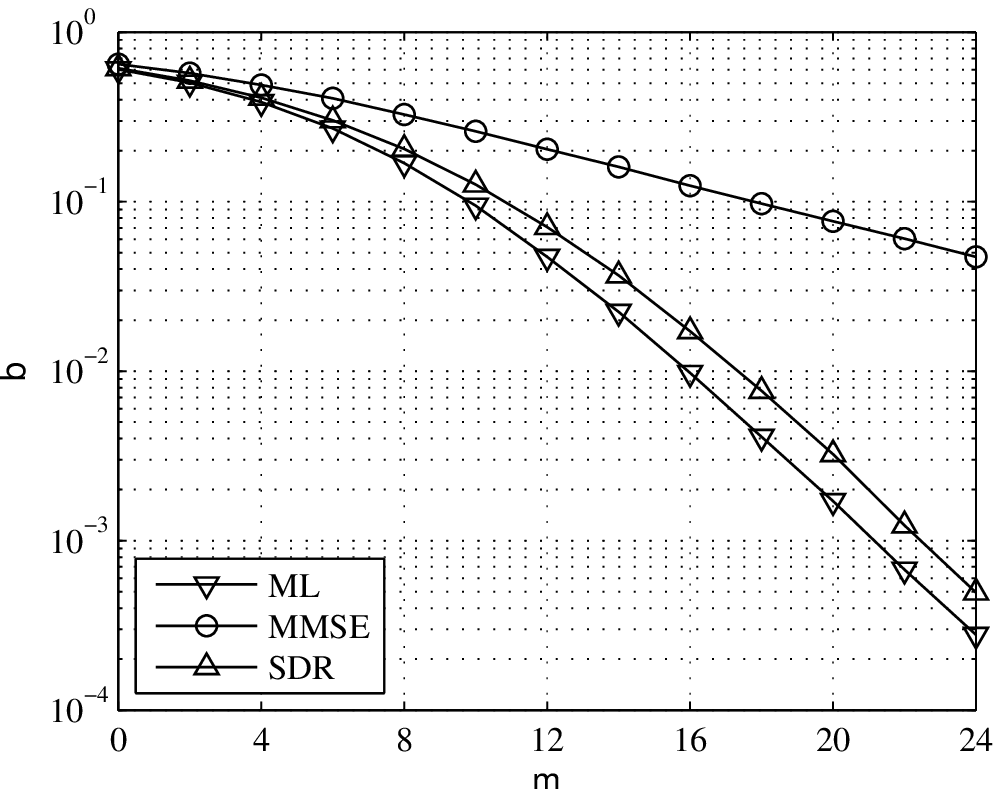}\\
\caption{The probability of error when $\bfH \in \reals^{n \times m}$ has i.i.d.~real valued Gaussian entries, and where $m = n = 4$.}\label{fig:realber}
\end{figure}

Thus, the use of suboptimal (but computationally advantageous) alternatives to ML detection is motivated. However, when applied to a fading channel there is unfortunately often a significant loss in performance associated with many of the suboptimal alternatives. This is illustrated in Fig.~\ref{fig:realber} where the probability of error for three different detectors is shown for the case where $\bfH \in \reals^{4 \times 4}$. By comparing the ML detector and minimum mean square error (MMSE) detector~\cite{TW:05} it can be seen that not only is the MMSE suboptimal, but the rate at which the probability of error tends to zero with increasing SNR is significantly lower than that of the optimal ML detector. This in turn results in a large loss in performance in the high SNR regime. The rate at which the error probability vanishes, or more precisely the slope (in log-log scale) of the error probability curve in the high SNR regime, is commonly referred to as the \emph{diversity} of the detector and it is well known that the MMSE detector has a significantly lower diversity than the ML detector~\cite{TW:05}. However, the third curve in Fig.~\ref{fig:realber} shows the probability of error for a receiver structure known as the semidefinite relaxation (SDR) detector or receiver. The SDR detector was (in the communications literature) first proposed in~\cite{TR:01,MDW:02,ANJ:02} for CDMA multiuser detection but is applicable for the detection of binary signals transmitted over any MIMO channel on the form of~\eqref{eq:data-model}. The SDR receiver is based on a convex relaxation technique where the optimization in~\eqref{eq:ml-detector} is simplified by first expanding the feasible set and then applying a rounding procedure to obtain an approximate solution to~\eqref{eq:ml-detector}. Note that this statement is also true for the zero forcing (ZF) and MMSE receivers where an unconstrained least squares problem (a regularized least squares problem in the MMSE case) is initially solved and where the symbol estimates are then obtained by componentwise threshold decisions. However, the semidefinite relaxation differs from ZF and MMSE receivers in that the problem is first lifted into a higher dimensional space before the relaxation takes place. From Fig.~\ref{fig:realber} it is apparent that the SDR receiver, although suboptimal in the sense that it does not achieve the minimum probability of error, does not suffer the loss in diversity experienced by the MMSE receiver.

The main contribution of this work is the analytic proof of the observation above. Namely, if the entries of $\bfH \in \reals^{n \times m}$ are i.i.d.~zero mean Gaussian with a finite variance and $n \geq m$, then the SDR receiver achieves the maximum possible receiver diversity. The result is formally stated in Theorem~\ref{th:main} in Section~\ref{sec:sdr-performance} and represents a non-trivial extension of previously known performance guarantees available for the SDR detector, see e.g.~\cite{Nes:97,MDW:02,KL:05}.

The topic of receiver diversity has received significant attention in the digital communications literature and other low complexity receivers have been designed specifically with diversity in mind. Perhaps, most prominent among these receivers are the lattice-reduction-aided (LRA) receivers~\cite{YW:02,WF:03}. In the LRA receiver one performs a change of basis under which the conditioning of $\bfH$ is improved and then applies a simple (e.g.~ZF, MMSE or decision feedback) detector in the new basis. It has also recently been shown that it is possible to construct (low complexity) full diversity receivers based on these ideas~\cite{TMK:06}, again under the assumption that $n \geq m$. However, the design philosophies underlying the LRA and SDR detectors are fundamentally different. Were as the LRA is combinatorial in nature the SDR detector is based on the minimization of a continuous function over a convex set. Further, in the LRA receiver it is assumed that the transmitted message belongs to an (infinite) integer lattice which enables the change of basis while in the SDR approach explicit use is made of the binary symbol assumption.

As previously stated, we treat the SDR receiver under the assumption that the channel matrix is i.i.d.~Gaussian and real valued. The main reason for this is that the SDR receiver is most easily treated in the real valued case. It should however be mentioned that the extension to the complex case is non-trivial and that numerical results suggest that a theorem, analogous to Theorem~\ref{th:main}, may not hold in this case. However, the numerical results also indicate that the loss in diversity (with respect to the ML detector) remains small. We discuss this issue further in Section~\ref{sec:complex}. Additionally, the underdetermined ($n < m$) case is treated in Section~\ref{sec:lowrank}. In the latter case our proof of Theorem~\ref{th:main} provides a lower bound on the diversity achieved by the SDR receiver which shows that if $m-n$ is not too large, then the diversity of the SDR is strictly larger than that of the MMSE and ZF receivers.

In Section~\ref{sec:sdr} we review the SDR receiver and present the main contribution of this work, namely Theorem~\ref{th:main}. In Section~\ref{sec:proof-outline} a short outline of the proof is given and the rigorous analysis is given in Section~\ref{sec:noisepart} and Section~\ref{sec:part2}. Further, a short discussion of how the results may possibly be generalized to other scenarios is given in Section~\ref{sec:estensions}. Also, although it makes no difference for the analytical results, we will in the numerical examples normalize the channel matrix, $\bfH$, such that each component has a variance of $n^{-1}$, yielding unit energy symbols at the receiver.

\section{Semidefinite Relaxation} \label{sec:sdr}

The use of semidefinite relaxation for bounding the optimal value of a combinatorial optimization problem was first considered in the late seventies~\cite{Lov:79} (where it was used to bound the Shannon capacity of a graph). Theoretical work in the nineties~\cite{LS:91} along with the introduction of practical methods for solving semidefinite programs~\cite{Jar:93,NN:94,VB:95} made the semidefinite relaxation a viable method for finding approximate solutions to many combinatorial problems. A famous example where the SDR technique can be applied is the \emph{max cut} problem in graph theory~\cite{GW:95}. The application of SDR to the detection problem considered herein has also been studied in the communications literature~\cite{TR:01,MDW:02,ANJ:02}.

We will in Section~\ref{sec:sdr-detector} provide a short review of the SDR detector in the communications context. It is not the intention to give a complete treatment of the SDR detector in terms of implementation or to discuss the various improvements which have been proposed but rather to introduce notation and capture specific assumptions made herein. The reader is instead referred to the original works~\cite{TR:01,MDW:02,ANJ:02} for a thorough treatment of the SDR detector in the context of digital communications. See also, apart from the above,~\cite{WSV:00} for a comprehensive collection of results regarding semidefinite programming in general and also specific results regarding the semidefinite relaxation technique.

\subsection{The SDR Detector} \label{sec:sdr-detector}

In order to introduce the semidefinite relaxation technique it is useful to note that the (non-convex) optimization problem given by
\begin{equation} \label{eq:unrelaxed}
\begin{array}{ll}
\underset{\bfX, \; \bfx}{\min} & \Trace(\bfL \bfX) \\
\text{s.t.} & \diag(\bfX) = \bfe \\
            & \bfX = \bfx \bfx\tr
\end{array}
\end{equation}
where $\bfe$ is the vector of all ones and where
\begin{equation} \label{eq:lmatrix}
\bfL \defeq \begin{bmatrix} \bfH\tr\bfH & -\bfH\tr\bfy \\
-\bfy\tr\bfH & \bfy\tr\bfy
\end{bmatrix}, \ \
\bfx \defeq \begin{bmatrix} \bfsh \\ 1 \end{bmatrix}
\end{equation}
is equivalent to~\eqref{eq:ml-detector} in the sense that the solution to~\eqref{eq:ml-detector} is easily obtained from the solution to~\eqref{eq:unrelaxed} and vice verse~\cite{TR:01,MDW:02,WSV:00}. Essentially, the formulation of~\eqref{eq:unrelaxed} is obtained by lifting~\eqref{eq:ml-detector} into a higher dimension where the criterion is linear in the optimization variable. The rank one constraint on~$\bfX$ along with the diagonal constraint ensure there is a one to one correspondence between the feasible sets of~\eqref{eq:ml-detector} and~\eqref{eq:unrelaxed}. The optimal point of~\eqref{eq:ml-detector} is related to the optimal point of ~\eqref{eq:unrelaxed} through $\bfx$ as shown in~\eqref{eq:lmatrix}.

As~\eqref{eq:unrelaxed} and~\eqref{eq:ml-detector} are equivalent they are also equally hard to solve from a complexity theoretic point of view. In particular, it follows from~\cite{Ver:89} that~\eqref{eq:relaxed} is also NP-hard in general. However, consider now instead the optimization problem given by
\begin{equation} \label{eq:relaxed}
\begin{array}{rl}
\underset{\bfX}{\min} & \Trace(\bfL \bfX) \\
\text{s.t.} & \diag(\bfX) = \bfe \\
            & \bfX \succeq \bfZero
\end{array}
\end{equation}
where $\bfX \succeq \bfZero$ means that $\bfX$ is symmetric and positive definite. Since $\bfX = \bfx\bfx\tr$ implies $\bfX \succeq \bfZero$ it follows that~\eqref{eq:relaxed} represents a relaxation of~\eqref{eq:unrelaxed}. The problem in~\eqref{eq:relaxed} is referred to as the semidefinite relaxation of~\eqref{eq:unrelaxed} (or equivalently~\eqref{eq:ml-detector}) and serves as the basis for the semidefinite relaxation detector.

It is useful to note that~\eqref{eq:relaxed} is a \emph{convex} problem which can be efficiently solved in polynomial time~\cite{NN:94,BV:04}. In particular, there is an interior point algorithm which solves~\eqref{eq:relaxed} to any fixed precision in $\bigO(m^{3.5})$ time~\cite{HRJ:96}, see also~\cite{TR:01} where this algorithm is presented in the digital communications context. In practice, only a few iterations with a complexity comparable to that of inverting an $m$ by $m$ matrix are required in order to obtain an approximate solution to~\eqref{eq:relaxed}.

It is straightforward to see that when the optimal solution to~\eqref{eq:relaxed} is rank one it is also an optimal solution to~\eqref{eq:unrelaxed}. The existence of rank one solutions to~\eqref{eq:relaxed} is however by no means guaranteed and in general, the solution to~\eqref{eq:relaxed} can only serve as a basis for obtaining an approximate solution to~\eqref{eq:unrelaxed}. In fact, it is possible to characterize exactly (in terms of $\bfH$, $\bfs$ and $\bfv$) when~\eqref{eq:relaxed} will and will not have rank one solutions, see~\cite{JMO:03} for necessary and sufficient conditions.

When the optimal point of~\eqref{eq:relaxed} is not rank one, some type of rounding procedure has to be used to round the optimal point of~\eqref{eq:relaxed} to a point in the feasible set~\eqref{eq:unrelaxed}. There are several suggestions for this in the literature. Among the more powerful approaches are a randomization technique~\cite{GW:95,MDW:02} and an approximation using the dominant eigenvector~\cite{TR:01}. Numerical evidence suggests that the randomization technique results in superior error performance. We shall however consider the very simple strategy of simply using the signs of the last column of $\bfX^\star$ where $\bfX^\star$ is an optimal point of~\eqref{eq:relaxed}. This approach was also mentioned in~\cite{TR:01} but discarded in favor of the (superior) eigenvector approach. However, as the simpler approach already achieves the maximum diversity we shall only consider this approach in detail. It should however be noted that the proof extends to the dominant eigenvector case in a straightforward manner by simply appealing to results regarding the continuity of eigenvectors corresponding to distinct (multiplicity one) eigenvalues.

To summarize, we obtain the SDR estimate, $\bfsh_\mathrm{SDR}$ as follows. Let $\bfX^\star$ be the minimizer of~\eqref{eq:relaxed}. Then $\bfsh_\mathrm{SDR}$ is defined according to
\begin{equation} \label{eq:rounding}
[\bfsh_\mathrm{SDR}]_i \defeq \mathrm{sgn}([\bfX^\star]_{i,m+1}), \quad i=1,\ldots,m
\end{equation}
where
$$
\mathrm{sgn}(x) = \left\{
\begin{array}{ll}
  1 & x > 0 \\
  -1 & x \leq 0\\
\end{array}
\right.
$$
is the sign function, i.e.~$\bfsh_\text{SDR}$ is given by the signs of the last column of $\bfX^\star$. Note that although it is possible for~\eqref{eq:relaxed} to have several optimal solutions it is always possible to pick some unique optimizer, $\bfX^\star$, from the optimal set. Thus, it can be assumed that $\bfsh_\text{SDR}$ is uniquely determined by $\bfy$ and $\bfH$.

Finally, it should be mentioned that extensions to the original semidefinite relaxation detectors have appeared in the literature. These include for example extensions to $M$-PSK constellations~\cite{MCD:04b} and $M$-QAM constellations~\cite{WES:05}. However, the analysis of these extensions is not treated herein.

\subsection{SDR Performance} \label{sec:sdr-performance}

The extraordinary performance of the SDR technique in many areas have been a motivating reason for its study and there are results in the literature regarding the quality of the semidefinite relaxation approximation of~\eqref{eq:unrelaxed} for more or less arbitrary choices of the matrix $\bfL$ (in~\eqref{eq:lmatrix}). These include the bound in~\cite{Nes:97} which is a generalization of a previous result for the max cut problem~\cite{GW:95}. There are also some results relating the semidefinite relaxation to other relaxations available for binary quadratic programs (such as~\eqref{eq:ml-detector})~\cite{PRW:95}.

In the context of digital communications it has previously been shown that several low complexity detectors may be viewed as further relaxations of the SDR detector~\cite{MDW:02}. Notably, these low complexity detectors include both the ZF and MMSE detectors and give strong support for the SDR approach although the results in~\cite{MDW:02} relate to the objective values of the relaxations rather than directly to the quality of the estimates, $\bfsh$. Further, a probabilistic bound on the difference in optimal objective value between~\eqref{eq:relaxed} and~\eqref{eq:unrelaxed} was given in~\cite{KL:05} for the large system limit. Also, as previously mentioned, the conditions for rank one solutions to~\eqref{eq:relaxed} were complectly characterized in~\cite{JMO:03} where it was also established that the detector was free of an error floor under the assumption that $\bfH\tr\bfH$ is full rank. However, the result in~\cite{JMO:03} does not extend to a statement regarding the diversity. Specifically, it is possible to show (using the result of~\cite{JMO:03}) that an alternative SDR receiver which calls an error whenever~\eqref{eq:relaxed} is not of rank one would not have the maximum diversity. In other words, the second phase of the SDR receiver where high rank solutions are used to obtain symbol estimates is crucial to the SDR performance and must be taken into account in the analysis.

The main contribution of this work is a rather strong statement regarding SDR performance when applied to a fading channel, namely that under the model in~\eqref{eq:data-model} with an i.i.d.~Gaussian channel for which $n \geq m$ the SDR detector will have a diversity equal to that of the optimal, ML, detector. Loosely speaking, although suboptimal, the SDR detector will have an error probability which vanishes at the same rate as the ML detector in the high SNR limit and the loss due to suboptimality will be a shift in SNR and not a loss of \emph{diversity}. We formally state this as follows.

\vspace{2mm}
\begin{theorem} \label{th:main}
Assume that $\bfH \in \reals^{n \times m}$ in~\eqref{eq:data-model} consist of i.i.d.~Gaussian entries of zero mean and fixed (non-zero) variance. Assume further that $n \geq m$. Then
$$
\lim_{\rho \rightarrow \infty} \frac{\ln \prob{\bfsh_\mathrm{SDR} \neq \bfs}}{\ln \rho} = \lim_{\rho \rightarrow \infty} \frac{\ln \prob{\bfsh_\mathrm{ML} \neq \bfs}}{\ln \rho} = -\frac{n}{2}.
$$
\end{theorem}
\vspace{3mm}

It is important to note that the SDR (and maximum) diversity is $\tfrac{n}{2}$ in this case and not $n$. This is because we explicitly consider a real valued channel matrix~\eqref{eq:data-model} as opposed to the complex channel case more frequently studied in the literature. It is straightforward to show the maximum achievable diversity in this case is $\tfrac{n}{2}$ by extending the proof of~\cite{NZA:00} to cover the real valued case. In the case of ZF and MMSE the diversity is $\tfrac{n-m+1}{2}$ which can be seen by following the argument of Section 8.5.1.~in~\cite{TW:05} with a real valued channel matrix.

Following~\cite{ZT:03} we will throughout this work make use of the symbol $\doteq$ to denote \emph{exponential equality}, defined according to
\begin{equation} \label{eq:doteq-def1}
f(\SNR) \doteq \SNR^{-d} \qquad \Leftrightarrow \qquad \lim_{\SNR \rightarrow \infty} \frac{\ln f(\SNR)}{\ln
\SNR} = -d.
\end{equation}
Similar definitions will also apply to the symbols $\dotleq$ and $\dotgeq$. For reference, we list the most important properties of the exponential equality in Appendix~\ref{app:expeq}. Using~\eqref{eq:doteq-def1} generally allows for a more compact (and suggestive) notation and in this notation the statement of Theorem~\ref{th:main} becomes
$$
\prob{\bfsh_\mathrm{SDR} \neq \bfs} \doteq \prob{\bfsh_\mathrm{ML} \neq \bfs} \doteq \rho^{-\frac{n}{2}}.
$$

Now, most of remaining part of this work is devoted to the proof of Theorem~\ref{th:main}. The formal proof is divided into several lemmas presented in Section~\ref{sec:noisepart} and Section~\ref{sec:part2}. However, before presenting the proof in full, a short outline is given in Section~\ref{sec:proof-outline}.

\section{The SDR Diversity Proof, Outline} \label{sec:proof-outline}

\begin{figure}
\begin{center}
\psfrag{H}{$\Hset$}
\psfrag{X}{$\Xset$}
\psfrag{Xe}{$\bfX_\bfe$}
\psfrag{x}[cc]{$x$}
\psfrag{y}[cc]{$y$}
\psfrag{z}[cc]{$z$}
\psfrag{1}[c]{\scriptsize $1$}
\psfrag{0}[c]{\scriptsize $0$}
\psfrag{-1}[c]{\scriptsize $-1$}
\includegraphics[width=\linewidth]{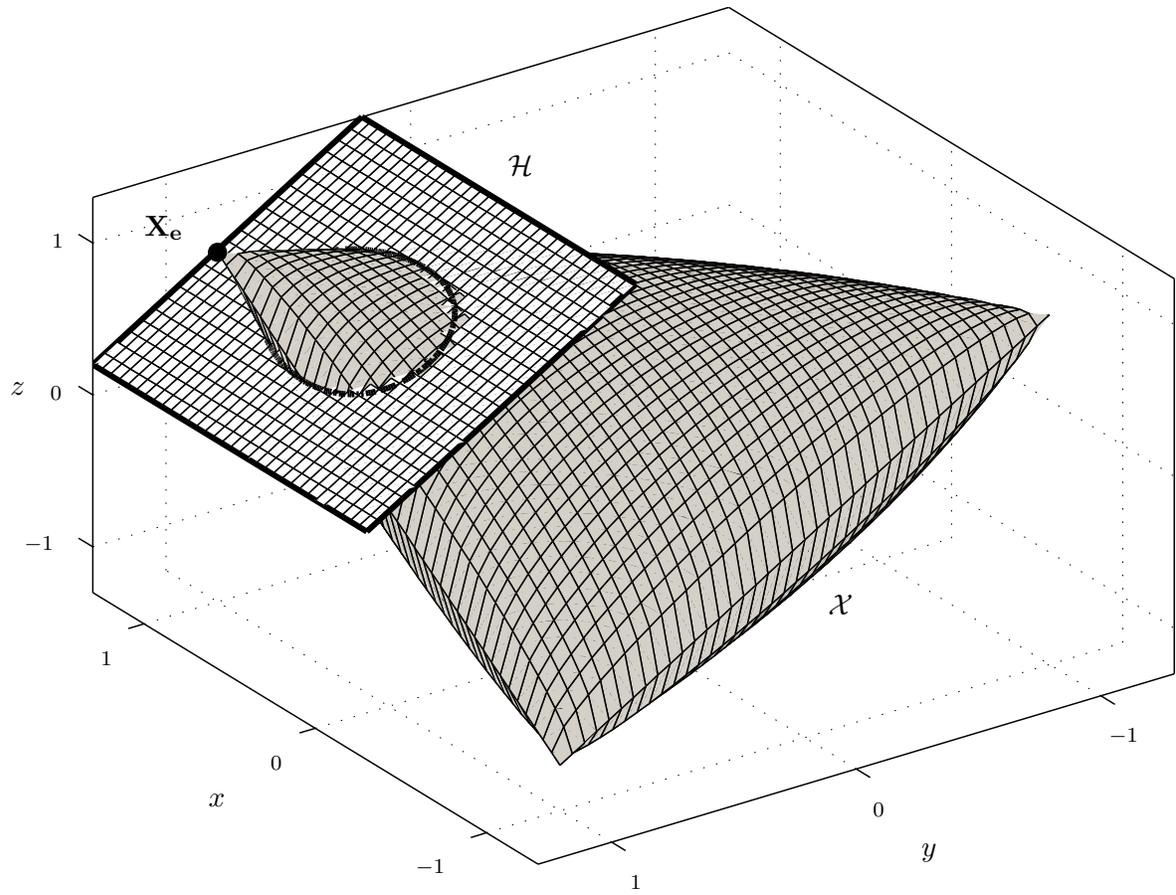}\\
\caption{Illustration of the feasible set, $\Xset$, of the SDR detector in~\eqref{eq:relaxed}. The hyperplane~$\Hset$ separates points in the feasible set that are close to and far from $\bfX_\bfe$.}\label{fig:hplane}
\end{center}
\end{figure}

Note that due to the symmetry of the problem (and the detector) it can without loss of generality be assumed that $\bfs = \bfe$ was transmitted. This will also be done in the sequel. In the $m=2$ case it is possible to graphically illustrate the feasible set, $\Xset$, of~\eqref{eq:relaxed} in order to gain intuition. To this end, consider parameterizing $\bfX \in \Xset$ as in~\cite{Hel:00} or~\cite{TR:01}, i.e.~according to
$$
\bfX = \begin{bmatrix}
  1 & x & y \\
  x & 1 & z \\
  y & z & 1 \\
\end{bmatrix}.
$$
The feasible set, $\Xset$, is illustrated in Fig.~\ref{fig:hplane}. The rank one matrix, $\bfX_\bfe$, that corresponds to the transmitted message, $\bfs = \bfe$, is also indicated in the figure.

Intuitively, one can characterize the error events of the SDR receiver as follows. When the optimal point of~\eqref{eq:relaxed}, $\bfX^\star$, is close to $\bfX_\bfe$ then the rounding procedure described in Section~\ref{sec:sdr} will be able to recover the correct rank one matrix, namely~$\bfX_\bfe$. It is only when the optimal point of~\eqref{eq:relaxed} is far from $\bfX_\bfe$ that an error can occur.

Consider now the introduction of a hyperplane, $\Hset$, as in Fig.~\ref{fig:hplane} that separates the points in $\Xset$ that are close to and far from $\bfX_\bfe$. Specifically, let $\Xset_{+}$ be the points in $\Xset$ that are on the same side of $\Hset$ as $\bfX_\bfe$ and let $\Xset_{-}$ be the points on the other side. Assume also that $\Hset$ is chosen such that points in $\Xset_{+}$ are rounded off to $\bfX_\bfe$. Let us also first consider the zero noise case, i.e.~when $\bfv=\bfZero$. In this case $\bfX_\bfe$ is always optimal for~\eqref{eq:relaxed} with a criterion value equal to 0. Further, let $\tau \geq 0$ be given by
$$
\tau = \min_{\bfX \in \Xset \cap \Hset} \Trace(\bfL\bfX),
$$
i.e.~$\tau$ is the minimum objective value over the intersection of the hyperplane and the feasible set, assuming $\bfv = \bfZero$. As the criterion function, $ \Trace(\bfL\bfX)$, is linear and $\Xset$ is convex it follows that the criterion function for any $\bfX \in \Xset_{-}$ will also satisfy $\Trace(\bfL\bfX) \geq \tau$.

Now allow for~$\bfv \neq \bfZero$ but assume that $\|\bfv\|$ is significantly smaller than $\tau$. In this case, $\Trace(\bfL\bfX_\bfe)$ is still small as $\Trace(\bfL\bfX_\bfe)$ is continuous in $\bfv$. At the same time it is guaranteed that $\Trace(\bfL\bfX)$ is not significantly smaller than $\tau$ for any $\bfX \in \Xset_{-}$, again since $\Trace(\bfL\bfX)$ is continuous in $\bfv$. This implies that there is a point in $\Xset_{+}$ with a criterion value close to zero, while all points in $\Xset_{-}$ have objective values which are at least on the order of $\tau$. In other words, the optimum over $\Xset$ must belong to $\Xset_{+}$ and therefore be close to $\bfX_\bfe$. This in turn implies that no error is made by the SDR receiver. In short, it is sufficient that $\tau$ is large in comparison with the noise in order for the detector to make a correct decision. This statement is also made rigorously by Lemma~\ref{lm:sufficient} in Section~\ref{sec:noisepart}.

The proof of Theorem~\ref{th:main} follows the heuristic argument given above and is divided into two parts. The first part, is concerned with proving that the error probability of the SDR detector is, in the high SNR regime, governed by the probability that $\tau$ is~\emph{atypically} small rather than the probability that $\bfv$ is atypically large. This statement is formalized by Lemma~\ref{lm:eq} in Section~\ref{sec:noisepart}. Note that the technique of interpreting typical errors as caused by particularly bad channels (in our case channels which cause $\tau$ to be small) is common in the literature, see e.g.~\cite{TW:05}. It is also similar in many respects to the analysis of coded multiple antenna systems where errors are typically caused by channels in \emph{outage}~\cite{ZT:03}.

The second part of the proof, contained in Section~\ref{sec:part2}, is concerned with bounding the probability that $\tau$ is atypically small. Note that in order for $\tau$ to be small there must be at least one $\bfX  \in \Xset \cap \Hset$ for which $\Trace(\bfL\bfX)$ is small. In essence, the technique used to establish our bound on the probability of $\tau$ being small can be summarized as follows.
\begin{enumerate}
\item Cover $\Xset \cap \Hset$ (or more precisely a set isomorphic to $\Xset \cap \Hset$) with $\epsilon$-balls and bound the probability that each specific $\epsilon$-ball contains an $\bfX$ for which $\Trace(\bfL\bfX)$ is small.
\item Count the number of $\epsilon$-balls required to cover $\Xset \cap \Hset$ and use the union bound to bound the probability that $\tau$ is small.
\end{enumerate}
Much of the difficulty of the proof stems from that the probability that each $\epsilon$-ball contains an $\bfX$ for which $\Trace(\bfL\bfX)$ is small depends on where in $\Xset \cap \Hset$ the $\epsilon$-ball is located. Also, the technically most challenging part of the proof relates to counting the number of $\epsilon$-balls required to cover certain subsets of $\Xset \cap \Hset$. The analysis of each particular $\epsilon$-ball is provided by Lemma~\ref{lm:ball} and the counting argument is captured in Lemma~\ref{lm:covering} in Section~\ref{sec:part2}. The proof of Theorem~\ref{th:main}, given at the end of Section~\ref{sec:part2}, then follows by combining Lemma~\ref{lm:ball} and Lemma~\ref{lm:covering}.

\section{The SDR Diversity Proof, Part I} \label{sec:noisepart}

The purpose of this section is to give rigorous justification of the first part of the heuristic argument given in Section~\ref{sec:proof-outline} and show that the noise, $\bfv$, can effectively be removed from (or integrated out of) the analysis of the receiver diversity. To this end, we will begin by giving a proper definition of some of the concepts appearing in the heuristic argument.

First of all, the feasible set, $\Xset$, of~\eqref{eq:relaxed} is given by
\begin{equation} \label{eq:Xset}
\Xset \defeq \{ \bfX \in \sym^{m+1} ~|~ \diag(\bfX) = \bfe,~\bfX \succeq \bfZero \}
\end{equation}
where $\sym^{m+1}$ denotes the set of symmetric matrices. Let $\Hset$ be the hyperplane (or affine subset of $\sym^{m+1}$) given by
\begin{equation} \label{eq:Hset}
\Hset \defeq \{ \bfX \in \sym^{m+1} ~|~ \Trace(\bfM\bfX\bfM\tr) = 1 \}
\end{equation}
where
\begin{equation} \label{eq:Mmatrix}
\bfM \defeq \begin{bmatrix}
  \bfI & -\bfe \\
\end{bmatrix} \in \reals^{m \times m+1}.
\end{equation}
It will later be established that an $\Hset$ chosen this way is sufficient for separating point close to $\bfX_\bfe$ from points far from $\bfX_\bfe$. The optimal value of $\Trace(\bfL\bfX)$ over the intersection set $\Xset \cap \Hset$ is under the zero noise, $\bfv = \bfZero$, assumption given by
\begin{equation} \label{eq:tau}
\tau \defeq \min_{\bfX \in \Xset \cap \Hset } \Trace(\bfL_\bfZero \bfX)
\end{equation}
where
$$
\bfL_\bfZero \defeq \begin{bmatrix}
  \bfQ & -\bfQ\bfe \\
  -\bfe\tr\bfQ & \bfe\tr\bfQ\bfe \\
\end{bmatrix} = \bfM\tr\bfQ\bfM
$$
and $\bfQ \defeq \bfH\tr\bfH$. Note that $\bfL_\bfZero$ is equal to $\bfL$ in~\eqref{eq:lmatrix} when $\bfv = \bfZero$. It is also straightforward to show that $\tau$ is equivalently given by
\begin{equation} \label{eq:tau2}
\tau = \inf_{\bfY \in \Yset} \Trace(\bfQ\bfY)
\end{equation}
where
\begin{equation} \label{eq:Yseta}
\Yset \defeq \bfM(\Xset \cap \Hset)\bfM\tr = \Ysett \cap \{\bfY \in \sym^m ~|~ \Trace(\bfY) = 1 \}
\end{equation}
and
\begin{equation} \label{eq:Ysett}
\Ysett = \bfM\Xset\bfM\tr.
\end{equation}
The set $\Ysett$ is a linear mapping of $\Xset \subset \sym^{m+1}$ onto $\sym^m$ given by $\bfM\bfX\bfM\tr$ under which the criterion $\Trace(\bfL_\bfZero\bfX)$ and $\Hset$ have a somewhat simpler structure. Note also that $\Ysett$ is convex since it is a linear transformation of a convex set. The main reason for introducing~\eqref{eq:tau2} is that it is frequently more convenient to work with~\eqref{eq:tau2} rather than with~\eqref{eq:tau} directly.

We are now able to pose and prove the first lemma regarding the error probability of the SDR detector. In essence, we wish to establish that a large $\tau$ is sufficient for correct detection. These statements are captured by Lemma~\ref{lm:sufficient} given below (note again that $\bfs = \bfe$ is assumed to be the transmitted message).

\vspace{1mm}
\begin{lemma} \label{lm:sufficient}
Let $\tau$ be given by~\eqref{eq:tau}. Then
$$
\tau > 4\|\bfv\|^2 \Rightarrow \bfsh_\mathrm{SDR} = \bfe.
$$
\end{lemma}
\vspace{2mm}

\noindent \emph{Proof:}
We will first prove the lemma under the assumption that the optimal point of ~\eqref{eq:relaxed} is rank deficient and then argue that this assumption can be made without loss of generality. Thus, consider an $\bfX \in \Xset$ for which $\bfX \nsucc \bfZero$ ($\bfX$ is positive semidefinite but not positive definite) and partition $\bfX$ as
$$
\bfX = \begin{bmatrix} \bfA\tr \\ \bfa\tr \\ \end{bmatrix}
\begin{bmatrix} \bfA & \bfa \\ \end{bmatrix} =
\begin{bmatrix}
  \bfA\tr\bfA & \bfA\tr\bfa \\
  \bfa\tr\bfA & \bfa\tr\bfa \\
\end{bmatrix}
$$
where $\bfA \in \reals^{m \times m}$ and $\bfa \in \reals^m$. Note that this is possible since $\bfX$ has at most rank $m$. Note also that $\|\bfa\|=1$ follows from $\diag(\bfX) = \bfe$. Further, note that the matrix $\bfL$ defined in~\eqref{eq:lmatrix} can be written as
$$
\bfL \defeq \begin{bmatrix}
  \bfH\tr\bfH & -\bfH\tr\bfy \\
  -\bfy\tr\bfH & \bfy\tr\bfy \\
\end{bmatrix} =
\begin{bmatrix} \bfH\tr \\ -\bfy\tr \\ \end{bmatrix}
\begin{bmatrix} \bfH & -\bfy \\ \end{bmatrix}.
$$
Thus,
\begin{align*}
\Trace(\bfL\bfX) = & \Trace\left(
\begin{bmatrix} \bfH\tr \\ -\bfy\tr \\ \end{bmatrix}
\begin{bmatrix} \bfH & -\bfy \\ \end{bmatrix}
\begin{bmatrix} \bfA\tr \\ \bfa\tr \\ \end{bmatrix}
\begin{bmatrix} \bfA & \bfa \\ \end{bmatrix}
\right) \\
= & \Trace\left(
\begin{bmatrix} \bfH & -\bfy \\ \end{bmatrix}
\begin{bmatrix} \bfA\tr \\ \bfa\tr \\ \end{bmatrix}
\begin{bmatrix} \bfA & \bfa \\ \end{bmatrix}
\begin{bmatrix} \bfH\tr \\ -\bfy\tr \\ \end{bmatrix}
\right) \\
= & \Trace((\bfH\bfA\tr - \bfy\bfa\tr)(\bfH\bfA\tr -
\bfy\bfa\tr)\tr) \\
= & \|\bfH\bfA\tr - \bfy\bfa\tr\|^2
\end{align*}
where $\|\cdot\|$ above refers to the the Frobenius norm. Now, the model of~\eqref{eq:data-model} for $\bfs = \bfe$ yields (through $\bfy$)
$$
\Trace(\bfL\bfX) = \|\bfH(\bfA\tr - \bfe\bfa\tr) -
\bfv\bfa\tr\|^2.
$$
Note that
\begin{align*}
& \| \bfH(\bfA-\bfe\bfa\tr) - \bfv\bfa\tr \| \\
\geq & \| \bfH(\bfA-\bfe\bfa\tr) \| - \|\bfv \bfa\tr\| \\
= & \|\bfH(\bfA-\bfe\bfa\tr) \| - \|\bfv\|
\end{align*}
where the last equality follows from $\|\bfa\| = 1$. Thus,
whenever
$$
\| \bfH(\bfA-\bfe\bfa\tr) \| > 2 \| \bfv \| \Leftrightarrow \| \bfH(\bfA-\bfe\bfa\tr) \|^2 > 4 \| \bfv \|^2
$$
it follows that
\begin{equation} \label{eq:trace-notoptimal}
\Trace(\bfL\bfX) > \|\bfv\|^2.
\end{equation}
At the same time, for
$$
\bfX_\bfe \defeq \begin{bmatrix}
  \bfe \\
  1 \\
\end{bmatrix}\begin{bmatrix}
  \bfe\tr & 1 \\
\end{bmatrix}
$$
it follows that
\begin{align} \label{eq:trace-optimal}
\Trace(\bfL\bfX_\bfe) = & \Trace\left(
\begin{bmatrix} \bfH\tr \\ -\bfy\tr \\ \end{bmatrix}
\begin{bmatrix} \bfH & -\bfy \\ \end{bmatrix}
\begin{bmatrix} \bfe \\ 1 \\ \end{bmatrix}
\begin{bmatrix} \bfe\tr & 1 \\ \end{bmatrix}
\right) \nonumber \\
= & \Trace\left(
\begin{bmatrix} \bfH & -\bfy \\ \end{bmatrix}
\begin{bmatrix} \bfe \\ 1 \\ \end{bmatrix}
\begin{bmatrix} \bfe\tr & 1 \\ \end{bmatrix}
\begin{bmatrix} \bfH\tr \\ -\bfy\tr \\ \end{bmatrix}
\right) \nonumber \\
= & \Trace((\bfH\bfe - \bfy)(\bfH\bfe - \bfy)\tr) \nonumber \\
= & \| \bfH\bfe - \bfy \|^2 = \| \bfv \|^2.
\end{align}
Thus, by~\eqref{eq:trace-notoptimal} and~\eqref{eq:trace-optimal}, it follows that
\begin{equation} \label{eq:noise-bound}
\| \bfH(\bfA-\bfe\bfa\tr) \|^2 > 4 \| \bfv \|^2 ~ \Rightarrow ~
\Trace(\bfL \bfX) > \Trace(\bfL \bfX_\bfe)
\end{equation}
which implies that $\bfX$ can not be optimal
for~\eqref{eq:relaxed} if
$$
\| \bfH(\bfA-\bfe\bfa\tr) \|^2 > 4 \| \bfv \|^2 \Leftrightarrow
\| \bfH(\bfA-\bfe\bfa\tr) \| > 2 \| \bfv \|.
$$
Now, note that
$$
(\bfA-\bfe\bfa\tr) = \bfM \begin{bmatrix} \bfA\tr \\ \bfa\tr \\
\end{bmatrix},
$$
for $\bfM$ defined in~\eqref{eq:Mmatrix} and
\begin{align} \label{eq:qformequiv}
& \| \bfH(\bfA-\bfe\bfa\tr) \|^2 \nonumber \\
= & \Trace\left( \bfH \bfM
\begin{bmatrix} \bfA\tr \\ \bfa\tr \\ \end{bmatrix}
\begin{bmatrix} \bfA & \bfa \\ \end{bmatrix}
 \bfM\tr \bfH\tr \right) \nonumber \\
= & \Trace \left( \bfH\tr \bfH \bfM
\begin{bmatrix} \bfA\tr \\ \bfa\tr \\ \end{bmatrix}
\begin{bmatrix} \bfA & \bfa \\ \end{bmatrix}
 \bfM\tr \right) \nonumber \\
= & \Trace(\bfH\tr\bfH\bfM\bfX\bfM\tr).
\end{align}
Let $\bfX^\star \in \Xset$ be the optimal point for~\eqref{eq:relaxed}
and let $\bfY^\star \in \Ysett$ be given by $\bfY^\star \defeq
\bfM\bfX^\star\bfM\tr$. Note that
$$
\Trace(\bfQ\bfY^\star) \leq 4\|\bfv\|^2
$$
for $\bfQ = \bfH\tr\bfH$ as otherwise $\bfX^\star$ would not be
optimal due to~\eqref{eq:noise-bound} and~\eqref{eq:qformequiv}.

Assume (as in the lemma) that
$$
\tau > 4\|\bfv\|^2.
$$
This implies that $\Trace(\bfQ\bfY) > 4\|\bfv\|^2$ for any $\bfY
\in \Yset$. The same conclusion could also be drawn for any
$\bfY \in \Ysett$ which satisfies $\Trace(\bfY) \geq 1$. This
follows since $\Ysett$ is a convex set which contains $\bfZero$ (since $\bfZero = \bfM\bfX_\bfe\bfM\tr$). That
is, if there were $\bfY \in \Ysett$ for which $\Trace(\bfY) \geq
1$ and $\Trace(\bfQ\bfY) \leq 4\|\bfv\|^2$ then $\bfYt
\defeq \gamma\bfY \in \Yset$ for some $\gamma \in (0,1]$ and
$\Trace(\bfQ\bfYt) \leq 4\|\bfv\|^2$ contrary to the assumption.

Thus, under the assumption of the lemma, it follows that
$$
\Trace(\bfY^\star) < 1
$$
and $\|\diag(\bfY^\star)\|_\infty < 1$ as $\bfY^\star \succeq \bfZero$ implies that $\bfY^\star$ has positive diagonal elements. Now, partition $\bfX^\star$ as
$$
\bfX^\star = \begin{bmatrix}
  \bfB & \bfb \\
  \bfb\tr & 1 \\
\end{bmatrix}
$$
where $\diag(\bfB) = \bfe$ due to~$\diag(\bfX^\star) = \bfe$. Computing $\bfY^\star$ explicitly under this partitioning yields
$$
\bfY^\star = \bfM\bfX^\star\bfM\tr = \bfB - \bfe\bfb\tr-\bfb\bfe\tr + \bfe\bfe\tr
$$
which implies
$$
\|\bfe - \bfb\|_\infty = \tfrac{1}{2} \|\diag(\bfY^\star)\|_\infty < \tfrac{1}{2}
$$
since $\diag(\bfY^\star) = 2\bfe - 2\bfb$. Thus, the rounding procedure given in~\eqref{eq:rounding} will round the last column of $\bfX^\star$, namely $\bfb$, to $\bfe$ and it follows that $\bfsh_\mathrm{SDR} = \bfe$.

What remains now is to show that the optimal point of~\eqref{eq:relaxed} must be rank deficient. By applying the result in~\cite{Pat:98} it is known that there will always be a rank deficient optimal point. A potential problem could arise if there are several optimal points, some of which are full rank. We will however show this that this is not possible.

In order for any optimal point of~\eqref{eq:relaxed} to be full rank, all off diagonal elements of $\bfL$ in~\eqref{eq:lmatrix} must be identically zero. This follows since otherwise there would be a search direction in the nullspace of $\diag(\bfX) = \bfe$ for which the criterion function would decrease, contradicting the optimality of any full rank $\bfX$. Thus $\bfH\tr\bfH$ has zero off diagonal elements (as it appears in $\bfL$) and $\bfH$ has orthogonal columns. In this special case the SDR will always have rank one solutions which are unique as long as the ML problem has a unique solution~\cite{JMO:03}. However, the assumption that $\tau > 4\|\bfv\|^2$ implies that
$$
\|\bfy-\bfH\bfe\|^2 < \|\bfy - \bfH\bfsh\|^2
$$
for any $\bfsh \in \bins^m$, $\bfsh \neq \bfe$, and it follows that the ML solution is unique. Therefore, there are no full rank solutions under the assumption in the lemma. This completes the proof. \hfill $\blacksquare$

Essentially, Lemma~\ref{lm:sufficient} states that for an error to occur in the high SNR regime one of two thing must happen. Either $\tau$ is atypically small or $\bfv$ is atypically large. As stated in Section~\ref{sec:proof-outline} it can be argued that the probability of the former event outweighs the probability of the latter. This is formally stated by the following Lemma which concludes this section.

\vspace{2mm}
\begin{lemma} \label{lm:eq}
Let $\tau$  be given by~\eqref{eq:tau}. Then
\begin{equation}
\prob{\tau \leq \rho^{-1}} \dotleq \rho^{-d} \qquad \Rightarrow \qquad \prob{\bfsh_\mathrm{SDR} \neq \bfe} \dotleq \SNR^{-d}.
\end{equation}
\end{lemma}
\vspace{2mm}

\noindent \emph{Proof:}
Assume (as was done in the lemma) that
$$
\prob{\tau \leq \rho^{-1}} \dotleq \rho^{-d}.
$$
This, combined with $\prob{\tau \leq \rho^{-1}} \leq 1$, implies that for any arbitrarily small $\delta > 0$ there is
a constant, $c$, for which
$$
\prob{\tau \leq \rho^{-1}} \leq c \rho^{-d+\delta}
$$
for all $\rho \geq 0$. Now, by Lemma~\ref{lm:sufficient},
$$
p_e \defeq \prob{\bfsh \neq \bfe} \leq \prob{\tau \leq
4\|\bfv\|^2}.
$$
Introduce a Gaussian vector, $\bfw \in \reals^n$, with i.i.d. zero mean elements of variance one and note that $\SNR^{-1}\|\bfw\|^2$ has the same distribution as $\|\bfv\|^2$. Let $f_{\|\bfw\|^2}(\gamma)$ denote the probability density function of $\gamma = \|\bfw\|^2$. As $\tau$ is independent of $\bfv$ (and $\bfw$) it follows that,
\begin{align*}
p_e \leq &~ \prob{\tau \leq 4\SNR^{-1}\|\bfw\|^2} \\
= & \int_0^\infty \prob{\tau \leq 4\SNR^{-1}\|\bfw\|^2~|~\|\bfw\|^2 = \gamma}f_{\|\bfw\|^2}(\gamma) d\gamma \\
= & \int_0^\infty \prob{\tau \leq 4\SNR^{-1}\gamma} f_{\|\bfw\|^2}(\gamma) d\gamma \\
\leq &~ c 4^{d-\delta} \SNR^{-d+\delta} \int_0^\infty \gamma^{d-\delta} f_{\|\bfw\|^2}(\gamma) d\gamma \\
= &~ c 4^{d-\delta} \SNR^{-d+\delta}
\expt{\|\bfw\|^{2(d-\delta)} } = c' \SNR^{-d+\delta}
\end{align*}
for some $c'$ independent of $\SNR$. Note that $c' < \infty$ follows since $\|\bfw\|$ has finite moments. Thus,
$$
p_e \dotleq \SNR^{-d+\delta}.
$$
However, as the relation holds for arbitrary small $\delta > 0$
it follows that
$$
p_e \dotleq \SNR^{-d}
$$
which concludes the proof.\hfill$\blacksquare$

\section{The SDR Diversity Proof, Part II} \label{sec:part2}

Let $\tau$ be given by~\eqref{eq:tau} or equivalently~\eqref{eq:tau2}. In light of Lemma~\ref{lm:eq} all that remains to be done in order to prove Theorem~\ref{th:main} is to provide a bound on
$$
\prob{\tau \leq \rho^{-1}}
$$
in the high SNR limit. Note however that at this point the variable $\rho^{-1}$ is just a dummy variable and we can, and will, replace $\rho^{-1}$ by $\epsilon$ and study the probability that $\tau \leq \epsilon$ for small $\epsilon > 0$. Thus, what remains to be done is to bound $\prob{\tau \leq \epsilon}$ around $\epsilon = 0$. We will also in the remaining part of this work focus on the optimization problem given in~\eqref{eq:tau2} rather than the equivalent problem in~\eqref{eq:tau}.

The probability that $\Trace(\bfQ\bfY) \leq \epsilon$ for some particular $\bfY \in \Yset$ will generally depend on the specific $\bfY$ considered (as mentioned in Section~\ref{sec:proof-outline}). In order to deal with this we shall first partition $\Yset$ into a finite number of subsets $\{ \Yset_i \}$,
$$
\Yset \subset \bigcup_i \Yset_i,
$$
such that $\prob{\Trace(\bfQ\bfY) \leq \epsilon}$ is more or less constant for all $\bfY$ within one such subset. Then, the probability that $\tau \leq \epsilon$ will be bounded by applying the union bound according to
\begin{equation} \label{eq:probsum}
\prob{\tau \leq \epsilon} \leq \sum_i \prob{\tau_i \leq \epsilon}
\end{equation}
where
$$
\tau_i \defeq \inf_{\bfY \in \Yset_i} \Trace(\bfQ\bfY)
$$
and where by property~\eqref{eq:doteq-sum} in Appendix~\ref{app:expeq} it is known that the sum in~\eqref{eq:probsum} will in the exponential equality sense be given (or completely dominated) by its maximal term.

It is interesting to note that this corresponds to the identification of \emph{typical} error events (or classes of error events), which is closely related to the analysis of typical \emph{outage} events in~\cite{ZT:03}. However, in~\cite{ZT:03} typical events where identified by classifying particularly bad channels, $\bfH$, while here, we shall use the concept to identify particularly troublesome subsets of $\Yset$. In essence, we shall partition $\Yset$ based on the eigenvalues of $\bfY \in \Yset$ (or how close to singular $\bfY$ is). Then the subset which dominates~\eqref{eq:probsum} will be found by optimizing over the possible eigenvalue combinations. Note also that these subsets will generally depend on $\epsilon$ but that we will adopt a somewhat casual terminology and refer to them simply as subsets rather than by the technically more correct term ``\emph{sequence} of subsets''. However, before considering the general partitioning of $\Yset$ into such subsets we will treat two motivating, and relatively simple, special cases to gain intuition.

\subsection{Special cases}

\subsubsection{Rank one matrices} \label{sssec:rankonediv}
First, let us consider the set of rank one matrices $\bfY \in \Yset$,~i.e. the set given by
$$
\Yset_\mathrm{R1} \defeq \Yset \cap \{ \bfY ~|~ \rank(\bfY) = 1 \}.
$$
For any particular $\bfY$ in this set, with an eigenvalue decomposition given by $\bfY = \sigma \bfu\bfu\tr$ where $\|\bfu\| = 1$, we have
\begin{equation} \label{eq:rankonediv}
\Trace(\bfQ\bfY) = \sigma \bfu\tr\bfQ\bfu.
\end{equation}
As $\sigma = 1$ due to the constraint $\Trace(\bfY) = 1$ it follows that
$$
\prob{ \Trace(\bfQ\bfY) \leq \epsilon } = \prob{\|\bfH\bfu\|^2 \leq \epsilon} \doteq \epsilon^\frac{n}{2}
$$
for this particular $\bfY \in \Yset_\mathrm{R1}$. It can also be shown that there are exactly $2^m-1$ distinct $\bfY \in \Yset_\mathrm{R1}$. In essence, each such $\bfY$ corresponds to the point at which line (in $\Xset$) connecting
$$
\bfX_\bfsh \defeq \begin{bmatrix}
  \bfsh \\
  1 \\
\end{bmatrix}\begin{bmatrix}
  \bfsh\tr & 1 \\
\end{bmatrix}
$$
and $\bfX_\bfe$ intersects the hyperplane $\Hset$, given in~\eqref{eq:Hset}. Therefore, by applying the union bound to the finite number of rank one $\bfY \in \Yset_\mathrm{R1}$ it follows that
$$
\prob{ \tau_\mathrm{R1} \leq \epsilon } \doteq \epsilon^\frac{n}{2}
$$
where
$$
\tau_\mathrm{R1} = \inf_{\bfY \in \Yset_\mathrm{R1}} \Trace(\bfQ\bfY).
$$
Note also that there is a one-to-one correspondence between the rank one matrices and all possible messages (not equal to the transmitted message), $\bfsh \in \bins^m \backslash \bfe$, that are searched over by the ML detector. This is also the reason why
$$
\prob{ \tau_\mathrm{R1} \leq \epsilon } \doteq \prob{ \bfsh_\text{ML} = \bfe }.
$$

\subsubsection{Full rank matrices} \label{sssec:fullrankdiv}
Next, consider the set of full rank (or more precisely \emph{well conditioned}) $\bfY \in \Yset$ given by
$$
\Yset_\mathrm{FR} \defeq \Yset \cap \{ \bfY ~|~ \bfY \succeq c\bfI \}
$$
for some constant $c > 0$, and let
$$
\tau_\mathrm{FR} \defeq \inf_{\bfY \in \Yset_\text{FR}} \Trace(\bfQ\bfY).
$$
As the criterion function, $\Trace(\bfQ\bfY)$, may be bounded as
$$
\Trace(\bfQ\bfY) \geq c\Trace(\bfQ) = c\|\bfH\|^2
$$
for any $\bfY \in \Yset_\mathrm{FR}$ it follows directly that
$$
\prob{ \tau_\mathrm{FR} \leq \epsilon } \dotleq \epsilon^\frac{mn}{2}
$$
by applying property~\eqref{eq:doteq-gvec-small} in Appendix~\ref{app:expeq}. This result can also be strengthened to show that
$$
\prob{ \tau_\mathrm{FR} \leq \epsilon } \doteq \epsilon^\frac{mn}{2}.
$$

\subsubsection{Discussion} \label{sec:rankdefdiscussions}

The implication of the result in Sections~\ref{sssec:rankonediv} and~\ref{sssec:fullrankdiv} is that the event that $\tau \leq \epsilon$ is (in the limit) much less likely to be caused by one of the matrices in $\Yset_\mathrm{FR}$ than one of the matrices in $\Yset_\mathrm{R1}$. The probability of the former is on the order of $\epsilon^\frac{mn}{2}$ while the later is only $\epsilon^\frac{n}{2}$ and $\epsilon^\frac{mn}{2} \ll \epsilon^\frac{n}{2}$ when $\epsilon$ is small (provided $m > 1$). Thus, (in a very loose sense) the reason for the high diversity of the SDR detector is that the elements added in the relaxation (the ones in $\Yset_\mathrm{FR}$) are less likely to cause errors than the elements already present in the feasible set of the ML detection problem (the ones in $\Yset_\mathrm{R1}$).

The question which however remains to be answered is if there is some other set of $\bfY$, somewhere between the full rank and rank one matrices, which can cause $\tau \leq \epsilon$ to occur with a probability substantially larger than $\epsilon^\frac{n}{2}$. The answer to this question is somewhat surprisingly \emph{no} provided that $n \geq m$ (but \emph{yes} in some $n < m$ cases). In fact, most of the remaining part of the paper is concerned with the formal proof of this statement.

\subsection{The General Case}

In the general case we consider sets on the form given by
\begin{equation} \label{eq:Yset}
\Yset(\bfa,\bfb) \defeq \Yset \cap \{ \bfY ~|~ \epsilon^{a_k} \leq \sigma_k(\bfY) \leq \epsilon^{b_k} \}
\end{equation}
where $\bfa = (a_1,\ldots,a_m)$, $\bfb = (b_1,\ldots,b_m)$ and $\sigma_k(\bfY)$ denotes the $k$th eigenvalue of $\bfY$. For notational convenience we will also in~\eqref{eq:Yset} interpret $\epsilon^{a_k}$ as $0$ for $a_k = \infty$ in order to allow one or more eigenvalues to be identically equal to zero. We can without loss of generality assume that the eigenvalues are ordered and that $0 \leq a_1 \leq \ldots \leq a_m$, $0 = b_1 \leq \ldots \leq b_m$ and $b_k \leq a_k$ for $k=1,\ldots,m$. Note that the assumption that $b_1 = 0$ can be made since~\eqref{eq:Yset} would, due to the $\Trace(\bfY) = 1$ constraint of $\Yset$ in~\eqref{eq:Yseta}, be empty otherwise. Similarly to before we define
\begin{equation} \label{eq:tauab}
\tau(\bfa,\bfb) \defeq \inf_{\bfY \in \Yset(\bfa,\bfb)} \Trace(\bfQ\bfY).
\end{equation}

In what follows, a bound on the probability of $\tau(\bfa,\bfb) \leq \epsilon$ is obtained by first partitioning $\Yset(\bfa,\bfb)$ into even smaller sets (essentially $\epsilon$-balls) and then using the union bound to bound $\prob{\tau(\bfa,\bfb) \leq \epsilon}$. It will be more convenient to work with a square root factorization of $\bfY \in \Yset$ instead of with $\bfY$ directly. Thus, we define a function,
\begin{equation} \label{eq:sqrtfactors}
\varphi : \sym^{m}_+ \mapsto \reals^{m \times m}
\end{equation}
(where $\sym^{m}_+$ denotes the set of symmetric, positive semidefinite matrices) for which $\bfA = \varphi(\bfY)$ satisfies $\bfA = \bfU\bfSigma^\frac{1}{2}$ and where $\bfU \bfSigma \bfU\tr = \bfY$ is the eigenvalue decomposition of $\bfY$. That is, $\varphi$ provides square root factors of $\bfY$ which have orthogonal columns with norms equal to $\sqrt{\sigma_i}$. Let $\Aset(\bfa,\bfb)$ be given by
\begin{equation} \label{eq:Asetab}
\Aset(\bfa,\bfb) \defeq \varphi(\Yset(\bfa,\bfb)),
\end{equation}
i.e.~$\Aset(\bfa,\bfb)$ is the set of square root factors which can be obtained from $\bfY \in \Yset(\bfa,\bfb)$. Note that $\Trace(\bfQ\bfY) = \| \bfH \bfA \|^2$ since $\bfQ = \bfH\tr\bfH$ and $\bfA = \varphi(\bfY)$. The random variable $\tau(\bfa,\bfb)$, defined in~\eqref{eq:tauab}, can thus be equivalently defined by
\begin{equation} \label{eq:tauab2}
\tau(\bfa,\bfb) = \inf_{\bfA \in \Aset(\bfa,\bfb)} \|\bfH\bfA\|^2.
\end{equation}

We are now ready to provide the first lemma regarding the probability that $\|\bfH\bfAt\|^2 \leq \epsilon$ for any $\bfAt$ in an $\epsilon^\frac{1}{2}$-ball around a given center point $\bfA \in \Aset(\bfa,\bfb)$.

\vspace{1mm}
\begin{lemma} \label{lm:ball}
Consider $\bfA \in \Aset(\bfa,\bfb)$ and define
\begin{equation} \label{eq:balldef}
\Aset_{\epsilon}(\bfA) \defeq \{ \bfAt ~|~ \|\bfAt-\bfA\| \leq \epsilon^\frac{1}{2} \}.
\end{equation}
Further, let
\begin{equation} \label{eq:tauA}
\tau(\bfA) \defeq \inf_{\bfAt \in \Aset_\epsilon(\bfA)} \|\bfH \bfAt\|^2.
\end{equation}
Then,
$$
\prob{ \tau(\bfA) \leq \epsilon} \dotleq \epsilon^{\nu}
\qquad
\text{where}
\qquad
\nu \defeq \sum_{k=1}^m \frac{n(1-a_k)^+}{2}.
$$
and where $(\cdot)^+ = \max(0,\cdot)$.
\end{lemma}
\vspace{1mm}

\noindent \emph{Proof:} Note that, due to the rotational symmetry of the distribution of $\bfH$, it can without loss of generality be assumed that $\bfA$ is diagonal (and equal to $\bfSigma^\frac{1}{2}$ where $\bfSigma$ is a diagonal matrix containing the eigenvalues of $\bfY \in \Yset$ for which $\bfA = \varphi(\bfY)$).

Pick some $\delta > 0$ and consider the event that
\begin{equation} \label{eq:BallDiv-Hconst}
\|\bfH\| \leq \epsilon^{-\delta}
\end{equation}
and where at least one column of $\bfH$, $\bfh_k$, satisfies
\begin{equation} \label{eq:BallDiv-hconst}
\|\bfh_k\| \geq 2\epsilon^{\frac{1-a_k}{2}-\delta}.
\end{equation}
We will first show that this event implies that $\tau(\bfA) > \epsilon$ and next that the event fails to occur with a probability which is no larger (in the $\dotleq$ sense) than $\epsilon^{\nu-nm\delta}$. Hence
\begin{align*}
\prob{\tau(\bfA) \leq \epsilon} \leq & ~\prob{ \|\bfH\| \geq \epsilon^{-\delta} \cup \|\bfh_k\| < 2\epsilon^{\frac{1-a_k}{2}-\delta} \; \forall k} \\
\dotleq & ~\epsilon^{\nu-nm\delta}.
\end{align*}

Note first that~\eqref{eq:BallDiv-hconst} implies
$$
\|\bfh_k\sigma_k^{\frac{1}{2}} \| \geq 2\epsilon^{\frac{1}{2}-\delta}
$$
for at least one $k$ since $\sigma_k \geq \epsilon^{a_k}$. Note also that this implies
$$
\|\bfH\bfA\| = \|\bfH\bfSigma^\frac{1}{2}\| \geq 2\epsilon^{\frac{1}{2}-\delta}.
$$
Now, consider $\|\bfH\bfAt\|$ for any $\bfAt$ satisfying $\|\bfAt-\bfA \| \leq \epsilon^\frac{1}{2}$. Under the additional assumption of~\eqref{eq:BallDiv-Hconst} it follows that
\begin{align*}
\|\bfH\bfAt\| = & \|\bfH\bfA - \bfH(\bfA-\bfAt) \| \\
\geq & \|\bfH\bfA\| - \|\bfH(\bfA-\bfAt)\| \\
& \geq 2\epsilon^{\frac{1}{2}-\delta} - \epsilon^{\frac{1}{2}-\delta} \\
& = \epsilon^{\frac{1}{2}-\delta} > \epsilon^\frac{1}{2}
\end{align*}
where the last inequality holds whenever $\epsilon \leq 1$. Note also that $\|\bfH\bfAt\| > \epsilon^\frac{1}{2}$ implies $\|\bfH\bfAt\|^2 > \epsilon$. Therefore,~\eqref{eq:BallDiv-Hconst} and~\eqref{eq:BallDiv-hconst} implies that $\tau(\bfA) > \epsilon$.

Now, consider the probability that~\eqref{eq:BallDiv-hconst} fails to hold, e.g.~that
$$
\|\bfh_k\| < 2\epsilon^{\frac{1-a_k}{2}-\delta}
$$
for all $k=1,\ldots,m$. As the columns of $\bfH$ are independent this probability can be upper bounded as
\begin{align*}
& \prob{\|\bfh_k\| < 2\epsilon^{\frac{1-a_k}{2}-\delta} \; \forall k} \\
= & \prod_{k=1}^m \prob{ \|\bfh_k\| < 2\epsilon^{\frac{1-a_i}{2}-\delta} } \\
\dotleq & \prod_{k=1}^m \epsilon^\frac{n(1-a_k-2\delta)^+}{2}
\leq \epsilon^{\nu-nm\delta}
\end{align*}
where we have used
$$
\prob{\|\bfh\| \leq \epsilon^\frac{c}{2} } = \prob{\|\bfh\|^2 \leq \epsilon^c} \dotleq \epsilon^{\frac{nc^+}{2}}
$$
according to~\eqref{eq:doteq-gvec-small} in Appendix~\ref{app:expeq} with $\epsilon = \rho^{-1}$. The probability that~\eqref{eq:BallDiv-Hconst} fails to hold can be upper bounded as
$$
\prob{\|\bfH\| > \epsilon^{-\delta} } \dotleq \epsilon^\infty
$$
according to~\eqref{eq:doteq-gvec-large} in Appendix~\ref{app:expeq}. Therefore, by applying the union bound,
\begin{align*}
\prob{\tau(\bfA) \leq \epsilon} \leq & ~\prob{ \|\bfH\| \geq \epsilon^{-\delta} \cup \|\bfh_k\| < 2\epsilon^{\frac{1-a_k}{2}-\delta} \; \forall k} \\
\dotleq & ~\epsilon^{\nu-nm\delta} + \epsilon^\infty \dotleq \epsilon^{\nu-nm\delta}.
\end{align*}
However, as $\delta > 0$ was arbitrary it follows that
$$
\prob{ \tau(\bfA) \leq \epsilon } \dotleq \epsilon^\nu
$$
which concludes the proof. \hfill $\blacksquare$

The next lemma provides a bound on the number of $\epsilon^\frac{1}{2}$-balls (defined as in~\eqref{eq:balldef}) which are required to completely cover the set $\Aset(\bfa,\bfb)$. Lemma~\ref{lm:covering} is the technically most difficult result of this work and we discuss this lemma below but save the the stringent proof for Appendix~\ref{app:proof}.

\vspace{1mm}
\begin{lemma} \label{lm:covering}
Let $\Aset(\bfa,\bfb)$ and $\Aset_\epsilon(\bfA)$ be defined as in~\eqref{eq:Asetab} and~\eqref{eq:balldef}, respectively. Then there is a collection of points, $\mathfrak{A} = \{ \bfA_i \}$, for which
$$
\Aset(\bfa,\bfb) \subset \bigcup_{\bfA_i \in \mathfrak{A}} \Aset_\epsilon(\bfA_i)
$$
and
$$
| \mathfrak{A} | \dotleq \epsilon^{-\mu}
$$
where $|\mathfrak{A}|$ denotes the number of elements of $\mathfrak{A}$ and where
\begin{equation} \label{eq:xi}
\mu \defeq \sum_{k=2}^m \frac{(m-k+2)(1-b_k)^+}{2}.
\end{equation}
\end{lemma}
\vspace{1mm}

\noindent \emph{Proof:} Given in Appendix~\ref{app:proof}. \hfill $\blacksquare$
\vspace{1mm}

Essentially, the proof of Lemma~\ref{lm:covering} relies on a geometric argument based on the dimensionality of low rank subsets of $\Aset$. Specifically, as part of the proof of Lemma~\ref{lm:covering} it is shown that the set of rank $r$ matrices $\bfA \in \Aset$, i.e.
$$
\Aset_{\mathrm{R}r} \defeq \Aset \cap \{\bfA ~|~ \rank(\bfA) = r \},
$$
is part of a $d_r$-dimensional (smooth) manifold where
$$
d_r \defeq \sum_{k=2}^r (m-k+2), \quad r=2,\ldots,m
$$
and $d_1 \defeq 0$. The manifold containing $\Aset_{\mathrm{R}r}$ is locally diffeomorphic (having a one-to-one differentiable relation) with the $d_r$-dimensional unit cube in $\reals^{d_r}$ (this is a property of any smooth $d_r$-dimensional manifold~\cite{Mil:65} and not specific to $\Aset_{\mathrm{R}r}$). The volume, $V$, covered by one $d_r$-dimensional $\epsilon^\frac{1}{2}$-ball is on the order of
$$
V \doteq (\epsilon^\frac{1}{2})^{d_r} = \epsilon^\frac{d_r}{2}
$$
and therefore one needs on the order of
\begin{equation} \label{eq:number-of-balls}
N \doteq \frac{1}{V} \doteq \epsilon^\frac{-d_r}{2}
\end{equation}
such $\epsilon^\frac{1}{2}$-balls to cover the unit cube in~$\reals^{d_r}$. By exploiting that there is a differentiable (and therefore continuous) map between the unit cube and the manifold this result carries over to a covering of $\Aset_{\mathrm{R}r}$.

Thus, the set of rank $r$ matrices, $\Aset_{\mathrm{R}r}$, can be covered by a collection of points, $\mathfrak{A}_r$, satisfying
$$
|\mathfrak{A}_r| \dotleq \epsilon^{-\mu_r}
$$
where
$$
\mu_r = \frac{d_r}{2} = \sum_{k=2}^r \frac{(m-k+2)}{2}.
$$
Extending this line of reasoning from rank $r$ dimensional subsets, $\Aset_{\mathrm{R}r}$, to subsets which are close to being low rank in the sense that the singular values of $\bfA$ are bounded by powers of $\epsilon$ yields the result stated in Lemma~\ref{lm:covering}. Note also that this is similar to the discussion following Theorem 4 in~\cite{ZT:03}.

Now, Lemma~\ref{lm:ball} and Lemma~\ref{lm:covering} can be combined in order to bound the probability that $\Aset(\bfa,\bfb)$ contains an $\bfA$ for which $\|\bfH\bfA\|^2 \leq \epsilon$. Then, by optimizing over $\bfa$ and $\bfb$, one can find the set of the form of $\Aset(\bfa,\bfb)$ most likely to contain such an $\bfA$. It can also be argued that this set will dominate the probability of error in the high SNR regime. These ideas are captured by the following lemma.

\vspace{1mm}
\begin{lemma} \label{lm:lastlemma}
Let $\tau$ be defined as in~\eqref{eq:tau}. Then
$$
\prob{\tau \leq \epsilon} \dotleq \epsilon^\zeta
$$
where
\begin{equation} \label{eq:zeta}
\zeta \defeq \inf_{1 \geq c_2 \geq \ldots \geq c_m \geq 0}
\frac{n}{2} + \sum_{k=2}^m \frac{(n-m+k-2)c_k}{2}.
\end{equation}
\end{lemma}
\vspace{1mm}

\noindent \emph{Proof:} Consider picking some $\bfb = (b_1,\ldots,b_m)$
for which $b_1 = 0$ and $b_1 \leq b_2 \leq \ldots \leq b_m \leq 1$ and choose a $\delta > 0$. Let $\bfa = (a_1,\ldots,a_m)$ be given such that $a_1 = \delta$ and $a_k = b_k + \delta$ if $b_k + \delta \leq 1$ or $a_k = \infty$ otherwise for $k=2,\ldots,m$.

The probability that $\tau(\bfa,\bfb) \leq \epsilon$ where $\tau(\bfa,\bfb)$ is defined in~\eqref{eq:tauab} can be bounded, using the union bound according as
$$
\prob{\tau(\bfa,\bfb) \leq \epsilon} \leq \sum_{\bfA_i \in \mathfrak{A}} \prob{\tau(\bfA_i) \leq \epsilon}
$$
where $\mathfrak{A}$ is chosen according to Lemma~\ref{lm:covering} and where $\tau(\bfA_i)$ is given by~\eqref{eq:tauA}. Each term in the sum is upper bounded by
$$
\prob{\tau(\bfA_i) \leq \epsilon} \dotleq \epsilon^\nu
$$
where $\nu$ is given in Lemma~\ref{lm:ball}. The number of terms in the sum is upper bounded by
$$
|\mathfrak{A}| \dotleq \epsilon^{-\mu}
$$
where $\mu$ is given by~\eqref{eq:xi}. Thus, the probability that $\tau(\bfa,\bfb) \leq \epsilon$ is bounded as
$$
\prob{\tau(\bfa,\bfb) \leq \epsilon} \dotleq \epsilon^{\nu-\mu}
$$
where
\begin{align*}
\nu-\mu  = & \sum_{k=1}^m \frac{n(1-a_k)^+}{2} - \sum_{k=2}^m \frac{(m-k+2)(1-b_k)^+}{2} \\
\geq & \frac{n}{2} + \sum_{k=2}^m \frac{(n-m+k-2)(1-b_k)^+}{2} - \frac{mn\delta}{2} \\
\geq & \zeta - \frac{mn\delta}{2}
\end{align*}
and where the property
$$
(1-a_k)^+ \geq (1-b_k)^+ - \delta
$$
(for $a_k$ chosen as above) was used to establish the first inequality. The second inequality follows by the definition of $\zeta$ in~\eqref{eq:zeta} along with $b_k \geq 0$.

Now, let
$$
\Aset \defeq \varphi(\Yset)
$$
where $\varphi$ is given by~\eqref{eq:sqrtfactors}. Note that we can pick a finite set of $\bfb \in [0,1]^m$, $\mathfrak{B} = \{ \bfb_i \}$, such that
\begin{equation} \label{eq:finalcover}
\Aset \subset \bigcup_{\bfb \in \mathfrak{B}} \Aset(\bfa,\bfb)
\end{equation}
where $\bfa = \bfa(\bfb)$ according to the above. This follows since by specifying $\bfb = (b_1,\ldots,b_m)$ we include the matrices $\bfY \in \Yset$ for which the $k$th eigenvalue satisfies $\epsilon^{b_k+\delta} \leq \sigma_k \leq \epsilon^{b_k}$ if $b_k < 1$ and $\sigma_k \leq \epsilon$ if $b_k = 1$. Thus we can cover the entire range of $\sigma_k \in [0,1]$ with a finite number of $b_k \in [0,1]$. For the special case of $k = 1$ we know that $\sigma_1$ is bounded away from $0$ due to $\Trace(\bfY) = 1$ which implies that $\sigma_1 \in [\epsilon^\delta,1]$ for sufficiently small $\epsilon$ given that $\delta > 0$ which is why $b_1 = 0$ can be assumed without loss of generality.

Using the union bound it follows that
\begin{align*}
\prob{\tau \leq \epsilon}
\leq & \sum_{\bfb \in \mathfrak{B}} \prob{\tau(\bfa,\bfb) \leq \epsilon} \\
\dotleq & \epsilon^{\zeta - \frac{mn\delta}{2}}
\end{align*}
since each term in the sum satisfies
$$
\prob{\tau(\bfa,\bfb) \leq \epsilon} \dotleq \epsilon^{\zeta - \frac{mn\delta}{2}}
$$
and the number of terms is finite. However, as $\delta > 0$ was arbitrary it follows that
$$
\prob{\tau(\bfa,\bfb) \leq \epsilon} \dotleq \epsilon^\zeta
$$
which concludes the proof. \hfill $\blacksquare$
\vspace{1mm}

In light of Lemma~\ref{lm:lastlemma} the proof of Theorem~\ref{th:main} is now almost trivial. All that remains is to compute $\zeta$ in~\eqref{eq:zeta} and apply Lemma~\ref{lm:eq}. We give the proof below.

\vspace{1mm}
\noindent\emph{Proof (of Theorem~\ref{th:main}):}
For the case where $n \geq m$ all terms in the sum appearing in~\eqref{eq:zeta} are non negative. Thus, the minimum in~\eqref{eq:zeta} is achieved for $c_2=\ldots=c_m = 0$ and it follows that
$$
\zeta = \frac{n}{2}.
$$
This, combined with Lemma~\ref{lm:eq}, proves that
$$
\prob{\bfsh_\mathrm{SDR} \neq \bfe} \dotleq \rho^{-\frac{n}{2}}.
$$
Next, note that the error probability of the SDR receiver is lower bounded by
$$
\prob{\bfsh_\mathrm{SDR} \neq \bfe} \geq \prob{\bfsh_\mathrm{ML} \neq \bfe} \doteq \rho^{-\frac{n}{2}}
$$
since the ML detector achieves the minimum probability of error. It therefore follows that
$$
\prob{\bfsh_\mathrm{SDR} \neq \bfe} \doteq \prob{\bfsh_\mathrm{ML} \neq \bfe} \doteq \rho^{-\frac{n}{2}}.
$$
By noting again that $\bfs = \bfe$ can be assumed without loss of generality the statement of Theorem 1 follows. \hfill $\blacksquare$
\vspace{1mm}

\section{Extensions} \label{sec:estensions}

At this stage, only the case of real valued systems on the form of~\eqref{eq:data-model} have been considered. Also, for the proof of Theorem~\ref{th:main} it was assumed that $n \geq m$. In this section, we discuss the extensions which would follow by relaxing these constraints and some illustrative numerical examples are given.

\subsection{The $n < m$ case} \label{sec:lowrank}

As stated above, full diversity has so far been shown under the condition that $n \geq m$. However, a careful inspection of the proofs show that the only part which explicitly relies on this assumption is when it is argued that $c_2 = \ldots = c_m = 0$ is an optimal point for~\eqref{eq:zeta} in the $n \geq m$ case. However, nontrivial bounds on the diversity will follow whenever $\zeta$ in~\eqref{eq:zeta} is strictly positive. The following theorem provides a lower bound on the diversity for the case when $n < m$.

\vspace{1mm}
\begin{theorem} \label{th:lowrank}
Given the assumptions of Theorem~\ref{th:main} but for $r \defeq m - n > 0$, it holds that
$$
\lim_{\rho \rightarrow \infty} \frac{\ln \prob{\bfsh_\mathrm{SDR} \neq \bfs}}{\ln \rho} \leq -d
$$
where
\begin{equation} \label{eq:lowrank-d}
d = \frac{1}{2} \left( m - \frac{r(r+3)}{2} \right)
\end{equation}
\end{theorem}
\vspace{1mm}
\noindent \emph{Proof:} All that needs to be done in this case is to find the optimum in~\eqref{eq:zeta} and apply Lemma~\ref{lm:eq}. To this end, note that the optimum of~\eqref{eq:zeta} is achieved for $c_k = 1$ for all $k$ satisfying $$
n-m+k-2 < 0 \Leftrightarrow k \leq m-n+1
$$
and $c_k = 0$ for $k$ satisfying
$$
n-m+k-2 \geq 0 \Leftrightarrow k \geq m-n+2.
$$
The value of $\zeta$ in~\eqref{eq:zeta} is thus given as
$$
\zeta = \frac{n}{2} + \sum_{k=2}^{m-n+1} \frac{n-m+k-2}{2} = \frac{1}{2} \left( m - \frac{r(r+3)}{2} \right)
$$
This completes the proof. \hfill $\blacksquare$
\vspace{1mm}

\begin{figure}
\psfrag{b}{$\prob{\bfsh \neq \bfs}$}
\psfrag{m}{SNR [dB]}
\includegraphics[width=1.0\linewidth]{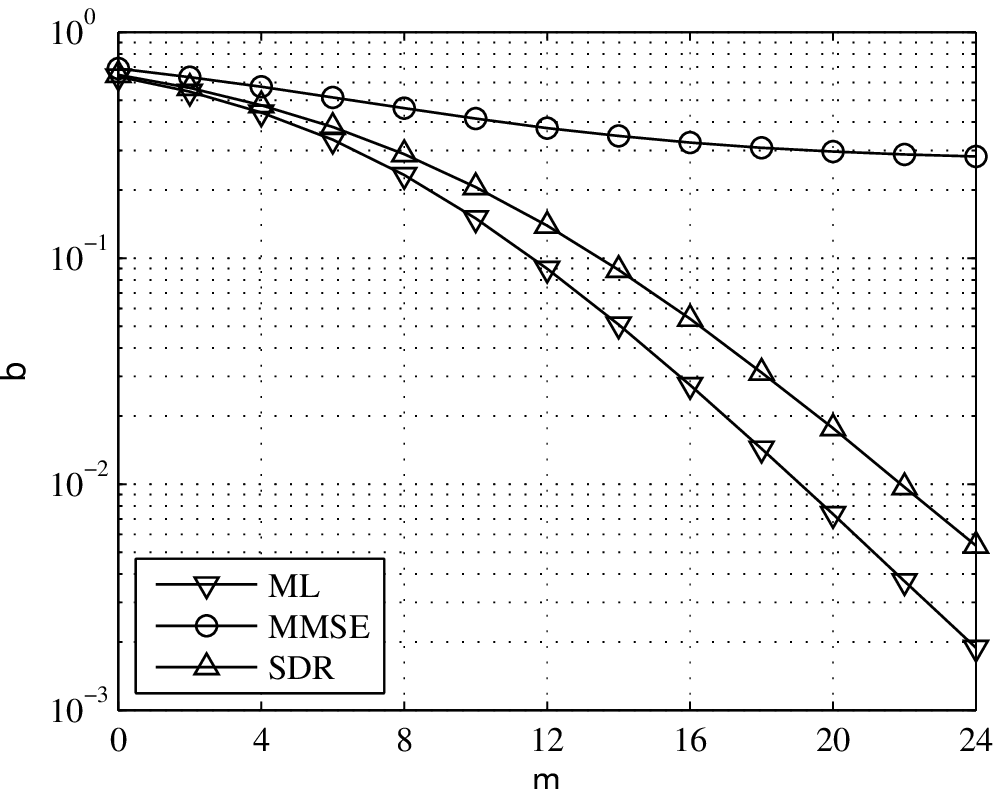}\\
\caption{The probability of error when $\bfH \in \reals^{n \times m}$ has i.i.d.~real valued Gaussian entries, and where $m = 4$ and $n = 3$.}\label{fig:realber3x4}
\end{figure}

\begin{figure}
\psfrag{b}{$\prob{\bfsh \neq \bfs}$}
\psfrag{m}{SNR [dB]}
\includegraphics[width=1.0\linewidth]{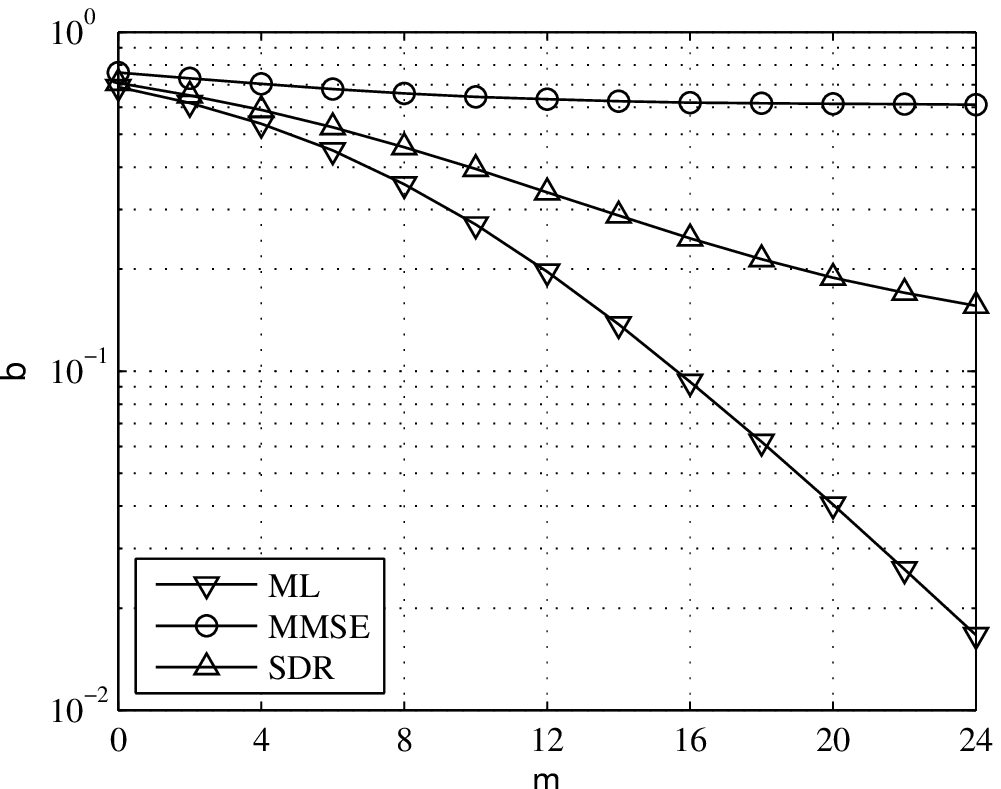}\\
\caption{The probability of error when $\bfH \in \reals^{n \times m}$ has i.i.d.~real valued Gaussian entries, and where $m = 4$ and $n = 2$.}\label{fig:realber2x4}
\end{figure}

Note that this result is only nontrivial if
$$
m > \frac{r(r+3)}{2}
$$
as otherwise Theorem~\ref{th:lowrank} would simply state that the probability of error is less than one. Further, we have no specific reason to believe that the bound is tight (in the sense that $\dotleq$ could be replaced by $\doteq$) in the $n < m$ case, even in the cases where the bound is non-trivial. An indication of this is given in Fig.~\ref{fig:realber3x4} where the diversity of the SDR detector seems to be larger than $2$ which is predicted by~\eqref{eq:lowrank-d}. It is however also unreasonable to expect the bound to be very loose in the sense that the SDR detector would maintain the same diversity as the ML detector in the general case where $n < m$. This is indicated by Fig.~\ref{fig:realber2x4} where the error probability of the SDR is significantly larger than that of the ML detector. Intuitively, in the $n < m$ case, it can become likely that a matrix with higher rank than one achieves the minimum in~\eqref{eq:tau2}. Therefore, the typical error events of the SDR detector no longer coincide with the error events of the ML detector and the SDR detector can experience a loss in diversity. We do not however, as pointed out above, expect the loss to be as large as what is indicated by~\eqref{eq:lowrank-d}.

A possible way to strengthen the analysis in the $n < m$ case can actually be seen by turning back to Fig.~\ref{fig:hplane}. Essentially, as part of proving Theorem~\ref{th:main} (and Theorem~\ref{th:lowrank}) the intersection of $\Xset$ and $\Hset$ is covered with $\epsilon$-balls. However, due to the linearity of the objective function it is already known that the minimum objective value over the intersection set must be achieved by one of the boundary points of $\Xset$. Therefore, it would suffice to cover the intersection of $\Hset$ with the \emph{boundary} of $\Xset$. This would in turn strengthen the bound on $|\mathfrak{A}|$ in Lemma~\ref{lm:covering} but would also require a framework for parameterizing the boundary set. It may also be possible to use the structure of the problem in other ways. One such way could be to make use of the results in~\cite{Pat:98} (where bounds on the rank of extremal matrices for semidefinite programs are provided) to further limit the part of the feasible set that needs to be covered.

\subsection{Complex channel matrices} \label{sec:complex}

It is well known that the SDR receiver is also applicable to the case where 4-QAM symbols are transmitted over a complex valued MIMO channel, see e.g.~\cite{TR:01}. The most direct strategy is to rewrite the problem in an equivalent real valued form according to
\begin{equation} \label{eq:data-model-complex}
\begin{bmatrix}
  \Re(\bfy_c) \\
  \Im(\bfy_c) \\
\end{bmatrix} = \begin{bmatrix}
  \Re(\bfH_c) & -\Im(\bfH_c) \\
  \Im(\bfH_c) & \Re(\bfH_c) \\
\end{bmatrix} \begin{bmatrix}
  \Re(\bfs_c) \\
  \Im(\bfs_c) \\
\end{bmatrix} + \begin{bmatrix}
  \Re(\bfv_c) \\
  \Im(\bfv_c) \\
\end{bmatrix}
\end{equation}
where $\bfy_c \in \complex^{N}$, $\bfH_c \in \complex^{N \times M}$, $\bfs_c \in \complex^M$ and $\bfv_c \in \complex^{N}$ are the (to~\eqref{eq:data-model}) corresponding complex valued quantities and where $\Re(\cdot)$ and $\Im(\cdot)$ denote the real and imaginary parts.

\begin{figure}
\psfrag{b}{$\prob{\bfsh \neq \bfs}$}
\psfrag{m}{SNR [dB]}
\includegraphics[width=1.0\linewidth]{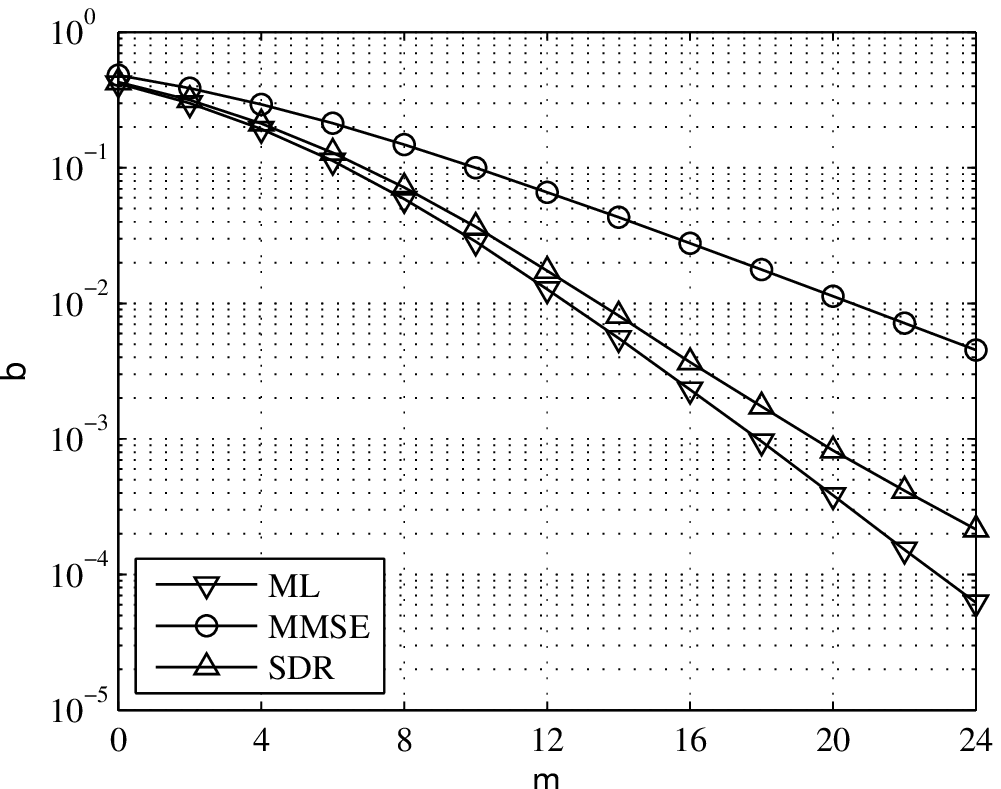}\\
\caption{The probability of error when $\bfH \in \complex^{N \times M}$ has i.i.d.~complex valued Gaussian entries, and where $N = M = 2$.}\label{fig:complexber2x2}
\end{figure}

\begin{figure}
\psfrag{b}{$\prob{\bfsh \neq \bfs}$}
\psfrag{m}{SNR [dB]}
\includegraphics[width=1.0\linewidth]{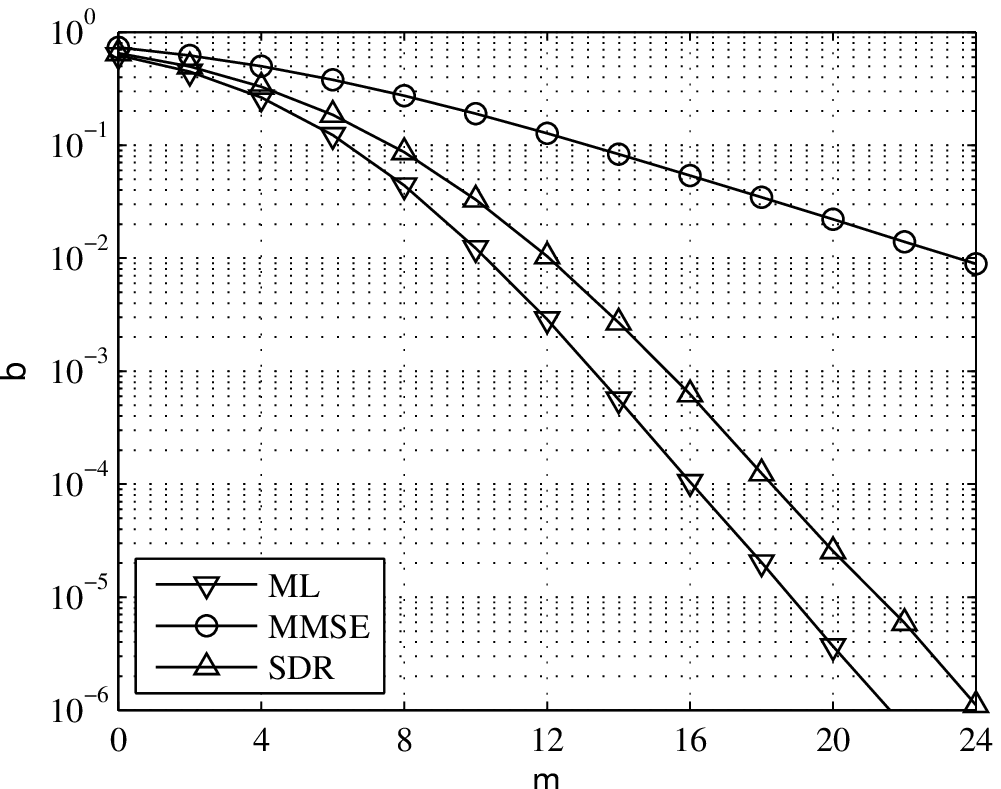}\\
\caption{The probability of error when $\bfH \in \complex^{N \times M}$ has i.i.d.~complex valued Gaussian entries, and where $N = M = 4$.}\label{fig:complexber4x4}
\end{figure}

However, the proof of Theorem~\ref{th:main} does unfortunately not extend to cover this case. The specific reason is found in Lemma~\ref{lm:ball} where the rotational symmetry of $\bfH$ is explicitly used. This symmetry is lost in the formulation given in~\eqref{eq:data-model-complex}, even in the case where $\bfH_c$ is i.i.d.~compex, circularly symmetric, zero mean Gaussian. More importantly, numerical simulations suggest that the extension of Theorem~\ref{th:main} to this case may not even be true. An indication of this can be seen in Fig.~\ref{fig:complexber2x2} where it is plausible to believe that the SDR receiver does experience a loss of diversity. However, it should also be pointed out that we do not expect the loss (if any) to be very large in general. This belief is based on extensive simulations, such as the one shown in Fig.~\ref{fig:complexber4x4}, that indicates a high SDR diversity in the complex case.

At first sight, what would be required in order to cover the complex case would be to update Lemma~\ref{lm:ball} for the structure of the effective channel matrix, $\bfH$, in~\eqref{eq:data-model-complex}. It is however also likely that Lemma~\ref{lm:covering} would need to be strengthened (as discussed in Section~\ref{sec:lowrank}) in order to obtain a tight bound on the diversity.  However, these steps remain a challenge. Also, note that if the SDR detector does not achieve full diversity, the issue of providing a lower bound on the error probability (or equivalently an upper bound on diversity) will also become more challenging.

\section{Conclusions}

In this paper we have shown that when applied to a fading channel, modelled by a real valued matrix with i.i.d.~Gaussian entries of zero mean and finite variance, the semidefinite relaxation detector achieves the maximum possible diversity. This provides a strong performance guarantee for the SDR approach, when applied in the communications context. Based on the discussions in Section~\ref{sec:estensions} it does not seem reasonable to expect such a strong statement to hold for an arbitrary system. Nonetheless, it is still reasonable to assume that the SDR detector will be superior to the class of linear detector and other relaxation techniques.

\begin{appendices}

\section{Exponential Equality} \label{app:expeq}

For the readers convenience, we list the (for this work) most important properties associated with the definition of \emph{exponential equality} in~\eqref{eq:doteq-def1}. These properties are easily derived from the definition in~\eqref{eq:doteq-def1} and can also be found (often implicitly) in many texts, see e.g.~\cite{ZT:03},~\cite{TW:05}. Thus, we state the properties without proof.

\begin{subequations}
\begin{enumerate}
\item \emph{Scaling property:}
For any $a \in [-\infty,\infty]$ and $c \in (-\infty,\infty)$ it holds that
\begin{equation} \label{eq:doteq-scaling}
f(\rho) \doteq \rho^{-a} \Rightarrow c f(\rho) \doteq \rho^{-a}.
\end{equation}

\item \emph{Summation property:}
For any $a,b \in [-\infty,\infty]$ it holds that
\begin{equation} \label{eq:doteq-sum}
f(\rho) \doteq \rho^{-a},~g(\rho) \doteq \rho^{-b} \Rightarrow f(\rho) + g(\rho) \doteq \rho^{-\min(a,b)}
\end{equation}
This property extends in the obvious way to the sum of finitely many terms.

\item \emph{Multiplication property:}
For any $a,b \in [-\infty,\infty]$ it holds that
\begin{equation} \label{eq:doteq-mult}
f(\rho) \doteq \rho^{-a},~g(\rho) \doteq \rho^{-b} \Rightarrow f(\rho)g(\rho) \doteq \rho^{-(a+b)}
\end{equation}
if the cases where $a+b$ is not well defined are excluded.

\item \emph{Extremal realizations of Gaussian vectors:}
Let $\bfh \in \reals^d$ be a vector of i.i.d. Gaussian elements of finite non-zero variance. Then
\begin{equation} \label{eq:doteq-gvec-small}
\prob{\|\bfh\|^2 \leq \rho^{-c}} \doteq \rho^{-\frac{d c^+}{2}}
\end{equation}
for $c \in (-\infty,\infty)$, where $c^+ \defeq \max(c,0)$ and
\begin{equation} \label{eq:doteq-gvec-large}
\prob{\|\bfh\|^2 \geq \rho^{c}} \doteq \rho^{-\infty}
\end{equation}
for $c > 0$. These properties follow by noting that $\|\bfh\|^2$ is $\chi^2$ distributed with $d$ degrees of freedom, see e.g.~\cite[Section 5.4.2]{TW:05}.
\end{enumerate}
\end{subequations}
It should also be noted that the properties given in~\eqref{eq:doteq-scaling}, \eqref{eq:doteq-sum} and \eqref{eq:doteq-mult} also hold with $\dotleq$ or $\dotgeq$ in place of $\doteq$.

\section{Proof of Lemma 4} \label{app:proof}

Before proving Lemma~\ref{lm:covering} we establish the following technical result regarding the feasible set of~\eqref{eq:tau2}.

\vspace{1mm}
\begin{lemma} \label{lm:changebasis}
The set $\Yset$ defined in~\eqref{eq:Yseta} satisfies
\begin{equation} \label{eq:Yseta2}
\Yset = \{ \bfY\in\sym^m ~|~ \Trace(\bfY) = 1, \bfY \succeq
\tfrac{1}{4}\bfd\bfd\tr,\bfd = \diag(\bfY) \}.
\end{equation}
\end{lemma}
\vspace{1mm}

\noindent \emph{Proof:} Consider the transformation given by
\begin{equation} \label{eq:Y-transform}
\underbrace{
\begin{bmatrix}
  \bfY & \bfa \\
  \bfa\tr & c \\
\end{bmatrix}}_{\bfP} =
\underbrace{
\begin{bmatrix}
  \bfI & -\bfe \\
  \bfZero\tr & 1 \\
\end{bmatrix}}_{\bfT} \bfX
\begin{bmatrix}
  \bfI & \bfZero \\
  -\bfe\tr & 1 \\
\end{bmatrix}
\end{equation}
or inversely,
\begin{equation} \label{eq:X-transform}
\bfX = \underbrace{
\begin{bmatrix}
  \bfI & \bfe \\
  \bfZero\tr & 1 \\
\end{bmatrix}}_{\bfR}
\begin{bmatrix}
  \bfY & \bfa \\
  \bfa\tr & c \\
\end{bmatrix}
\begin{bmatrix}
  \bfI & \bfZero \\
  \bfe\tr & 1 \\
\end{bmatrix}
\end{equation}
since $\bfT^{-1} = \bfR$. Note also that $\bfY$ is given by $\bfY =
\bfM\bfX\bfM\tr$ as $\bfM = \begin{bmatrix} \bfI & -\bfe
\\\end{bmatrix}$ by~\eqref{eq:Mmatrix}. Expanding $\bfX$
from~\eqref{eq:X-transform} yields
$$
\bfX = \begin{bmatrix}
  \bfY+\bfa\bfe\tr+\bfe\bfa\tr+\bfe c \bfe\tr& \bfa+\bfe c \\
  \bfa\tr+c\bfe\tr & c \\
\end{bmatrix}.
$$
Thus, the constraint $\diag(\bfX) = \bfe$ for $\bfX \in \Xset$ implies that $c=1$ for $\bfY \in \Yset$ since $\Yset \subset \Ysett = \bfM\Xset\bfM\tr$ for $\Yset$ given in~\eqref{eq:Yseta} and where $\Ysett$ is given in~\eqref{eq:Ysett}. Further, for $c=1$
$$
\diag(\bfY+\bfa\bfe\tr+\bfe\bfa\tr+\bfe \bfe\tr) = \diag(\bfY) +
2\bfa + \bfe = \bfe
$$
which implies that
\begin{equation} \label{eq:a-constraint}
\bfa = -\tfrac{1}{2} \diag(\bfY).
\end{equation}
Thus, given a matrix $\bfY \in \Ysett$ there is actually a unique $\bfX \in
\Xset$ for which $\bfY = \bfM\bfX\bfM\tr$. In other words, the
mapping from $\Xset$ to $\Ysett$ is one-to-one.

Since $\bfT$ (and $\bfR$) are invertible the constraint
$\bfX \succeq \bfZero$ is equivalent to $\bfP \succeq \bfZero$.
However, $\bfP \succeq \bfZero$ if and only if its Schur
complement~\cite{BV:04} is positive semidefinite, i.e.~if
$$
\bfY - c^{-1}\bfa\bfa\tr \succeq \bfZero.
$$
Thus, by combining~\eqref{eq:a-constraint} with $c = 1$ and identifying $\bfd = -2\bfa$ the equalities of~\eqref{eq:Yseta} and~\eqref{eq:Yseta2} are established. \hfill
$\blacksquare$

We are now in a position to prove the statement given by Lemma~\ref{lm:covering}. For convenience the lemma is restated below.

\vspace{1mm}
\emph{Lemma~\ref{lm:covering}:}~
Let $\Aset(\bfa,\bfb)$ and $\Aset_\epsilon(\bfA)$ be defined as in~\eqref{eq:Asetab} and~\eqref{eq:balldef} respectively. Then there is a collection of points, $\mathfrak{A} = \{ \bfA_i \}$, for which
$$
\Aset(\bfa,\bfb) \subset \bigcup_{\bfA_i \in \mathfrak{A}} \Aset_\epsilon(\bfA_i)
$$
and
$$
| \mathfrak{A} | \dotleq \epsilon^{-\mu}
$$
where
$$
\mu \defeq \sum_{k=2}^m \frac{(m-k+2)(1-b_k)^+}{2}.
$$

\emph{Proof:} Consider the triplet $(\bfU,\bflambda,\bfz) \in \reals^{m \times m}\times \reals^m\times \reals^m$ and the system of equations given by
\begin{subequations} \label{eq:svdset-eq}
\begin{align}
\Trace(\bfLambda^2) & = 1 \label{eq:svdset-trace} \\
\diag(\bfU\bfLambda^2\bfU\tr) &= \bfU\bfz \label{eq:svdset-range} \\
\bfU\tr\bfU &= \bfI \label{eq:svdset-orth} \\
\bfLambda^2 - \tfrac{1}{4} \bfz\bfz\tr & \succeq \bfZero \label{eq:svdset-bound}
\end{align}
\end{subequations}
where $\bfLambda \defeq \diag(\bflambda)$. The set of solutions to~\eqref{eq:svdset-eq} will in what follows be denoted by $\Mset$. The set of solutions to~\eqref{eq:svdset-trace},~\eqref{eq:svdset-range} and~\eqref{eq:svdset-orth} but not necessarily~\eqref{eq:svdset-bound} is denoted by $\Nset$ and it follows that $\Mset \subset \Nset$. From~\eqref{eq:svdset-trace} and~\eqref{eq:svdset-orth} it follows that $\bflambda$ and $\bfU$ in the solution set are bounded. However, as $\bfU$ is full rank due to~\eqref{eq:svdset-orth} it follows through~\eqref{eq:svdset-range} that $\bfz$ is also bounded. Therefore, both $\Nset$ and $\Mset$ are compact (closed and bounded) sets.

The constraints of~\eqref{eq:svdset-eq} are such that any solution, $(\bfU,\bflambda,\bfz)$, of~\eqref{eq:svdset-eq} satisfies $\bfU\bfLambda^2\bfU\tr \in \Yset$ and any eigenvalue decomposition, $\bfY = \bfU\bfSigma\bfU\tr$, of $\bfY \in \Yset$ solves~\eqref{eq:svdset-eq} for $\bfLambda = \bfSigma^\frac{1}{2}$ and some (unique) $\bfz$. To see this, consider the eigenvalue decomposition, $\bfY = \bfU\bfSigma\bfU\tr$, of some $\bfY \in \Yset$ where $\Yset$ is given by~\eqref{eq:Yseta}. Note also that $\bfY$ belongs to $\Yset$ if and only if it satisfies the constraints of~\eqref{eq:Yseta2} as proven in Lemma~\ref{lm:changebasis}. The orthogonality of $\bfU \in \reals^{m \times m}$ is a property of the eigenvalue decomposition and therefore~\eqref{eq:svdset-orth} is satisfied. For $\bfLambda=\bfSigma^\frac{1}{2}$ and $\bfz = \bfU\tr\diag(\bfU\bfLambda^2\bfU\tr)$ the constraint of~\eqref{eq:svdset-range} is satisfied. As $\bfY \in \Yset$ it follows that $\bfY - \frac{1}{4}\bfd\bfd\tr \succeq \bfZero$ where $\bfd = \diag(\bfY)$. Therefore, $\diag(\bfY) = \diag(\bfU\bfLambda^2\bfU\tr) = \bfU\bfz$ implies
$$
\bfU\bfLambda^2\bfU\tr -\tfrac{1}{4}\bfU\bfz\bfz\tr\bfU\tr \succeq \bfZero
\Leftrightarrow \bfLambda^2 - \tfrac{1}{4} \bfz\bfz\tr \succeq \bfZero
$$
which means that~\eqref{eq:svdset-bound} is satisfied. Finally, the constraint $\Trace(\bfY) = 1$ in~\eqref{eq:Yseta2} implies $\Trace(\bfLambda^2) = 1$ and~\eqref{eq:svdset-trace} is satisfied. Reversing the reasoning and applying Lemma~\ref{lm:changebasis} show that any solution to~\eqref{eq:svdset-eq} must also have the property that $\bfU\bfLambda^2\bfU\tr \in \Yset$.

The value of introducing~\eqref{eq:svdset-eq} is that it will, through the implicit function theorem~\cite{Rud:76}, provide a means of parameterizing the eigenvalues and vectors of $\bfY \in \Yset$. To this end, let
$$
p \defeq m+\frac{m(m+1)}{2}+1,
$$
$$
q \defeq m^2+2m,
$$
and $\bfomega \in \reals^{q}$ be given by
$$
\bfomega \defeq (\bfU,\bflambda,\bfz).
$$
Define
$$
H : \reals^q \mapsto \reals^p
$$
according to
$$
H(\bfomega) \defeq \begin{bmatrix}
\Trace(\bfLambda^2) - 1 \\
\diag(\bfU\bfLambda^2\bfU\tr) - \bfU\bfz \\
\svec(\bfU\tr\bfU-\bfI) \\
\end{bmatrix}
$$
and note that $H(\bfomega) = \bfZero$ corresponds to~\eqref{eq:svdset-trace},~\eqref{eq:svdset-range} and~\eqref{eq:svdset-orth}. In the above, $\svec(\cdot)$ referrers to the vector obtained by stacking the upper triangular part of a symmetric matrix into a vector. Let
$$
\bfomegab \defeq (\bfUb,\bflambdab,\bfzb)
$$
be a solution of~\eqref{eq:svdset-eq} and $\Iset$ be an index set satisfying
\begin{equation} \label{eq:Iset-sub}
\Iset \subset \left\{ 1,\ldots,q \right\}
\end{equation}
and
\begin{equation} \label{eq:Iset-size}
|\Iset| = p.
\end{equation}
Denote by $\bfomega_\Iset \in \reals^p$ the vector of components in $\bfomega$ indexed by $\Iset$ and let $\bfomega_{\Iset^c} \in \reals^{q-p}$ be the vector consisting of the remaining components. The implicit function theorem~\cite{Rud:76} states that if
\begin{equation} \label{eq:jacobian-criterion}
\left| \frac{\partial H(\bfomega)}{\partial \bfomega_\Iset}  \right|_{\bfomega=\bfomegab} \neq 0,
\end{equation}
then there is a neighborhood, $\Uset \subset \reals^q$, containing $\bfomegab$ and a differentiable mapping
$$
g : \reals^{q-p} \mapsto \reals^p
$$
satisfying $\bfomega_\Iset = g(\bfomega_{\Iset^c})$ for any $\bfomega \in \Uset \cap H^{-1}(\{\bfZero\})$.

Further~\eqref{eq:jacobian-criterion} implies the existence of a differentiable mapping
$$
\psi : \Dset \mapsto \Rset
$$
for which $\bfomega = \psi(\bfxi)$, where $\bfxi \defeq \bfomega_{\Iset^c} - \bfomegab_{\Iset^c} \in \reals^{q-p}$, where $\Dset$ is an open subset of $\reals^{q-p}$ containing $\bfZero$ and where $\Rset \defeq \psi(\Dset) \subset \reals^q$. This mapping is easily obtained from $g$ by including the components in $\bfomega_{\Iset^c}$ and performing a translation to a neighborhood of $\bfZero$. Thus, assuming that~\eqref{eq:jacobian-criterion} is satisfied, the solution set of~\eqref{eq:svdset-eq} is locally parameterized by $q-p$ scalar parameters. It will in fact later be shown that given \emph{any} solution, $\bfomegab$, to~\eqref{eq:svdset-eq} there will be some index set, $\Iset$, satisfying~\eqref{eq:Iset-sub} and~\eqref{eq:Iset-size} for which~\eqref{eq:jacobian-criterion} is satisfied. This implies that $\Nset$ is a $q-p$ dimensional (smooth) manifold embedded in $\reals^q$~\cite{Boo:86}. Note however that the specific index set, $\Iset$, required to satisfy~\eqref{eq:jacobian-criterion} will generally depend on the particular $\bfomegab$ chosen. This is analogous to the problem of parameterizing the unit circle based on solving $x^2+y^2=1$ where the choice of $x$ or $y$ as the \emph{free} parameter depends on if the parametrization neighborhood should include $x=0$ or $y=0$.

Note that it can without loss of generality be assumed that the domain of $\psi$, is given by
\begin{equation} \label{eq:neighborhood}
\Dset = (-\kappa,\kappa)^{q-p},
\end{equation}
i.e.~that $\Dset$ is an open hypercube for some $\kappa > 0$~\cite{Boo:86}. Further, since $\Nset$ is compact it can be assumed that $\kappa$ is independent of $\bfomegab$. It can also, without loss of generality, be assumed that $\psi$ is Lipschitz continuous~\cite{Bar:64} on $\Dset$. This follows since the inverse function theorem guarantees that $\psi$ has continuous derivatives on the closure of $\Dset$, $\bar{\Dset}$ (actually, in its standard form the inverse function theorem guarantees continuous derivatives on $\Dset$ but by reducing $\kappa$ if necessary the continuity can be extended to the closure of $\Dset$). Further, again due to the compactness of $\Nset$, it can be assumed that the Lipschitz constant of $\psi$ is independent of $\bfomegab$.

In order to prove the \emph{existence} of an index set, $\Iset$, for which~\eqref{eq:jacobian-criterion} is satisfied it is sufficient to prove that the Jacobian matrix $\bfD$,
\begin{equation} \label{eq:fulldiffmatrix}
\bfD \defeq \left. \frac{\partial H(\bfomega)}{\partial \bfomega} \right|_{\bfomega = \bfomegab} \in \reals^{p \times q},
\end{equation}
is full rank. In this event, the index set, $\Iset$, can be taken as the indexes of any $p$ linearly independent columns of~$\bfD$. For our purposes however, we shall need to be a bit more specific about how $\Iset$ is chosen. Therefore, note again that it will be of particular interest to study parameterizations of $\Mset$ (and $\Nset$) around solutions $\bfomegab$ corresponding to rank deficient $\bfY \in \Yset$ (see the discussion in Section~\ref{sec:rankdefdiscussions}). To this end, consider some $\bfomegab \in \Mset$ for which $\lambda_{r+1} = \ldots = \lambda_m = 0$, i.e.~$\bfomegab$ corresponds to a rank $r$ matrix $\bfYb \in \Yset$. Here, and in what follows, $\lambda_k$ and $z_k$ refer to the $k$th component of $\bflambda$ and $\bfz$ respectively. For any $\bfomegab \in \Mset$ it follows by~\eqref{eq:svdset-bound} that $|z_k| \leq 2|\lambda_k|$ for $k=1,\ldots,m$ and in particular it follows that $z_k = 0$ whenever $\lambda_k = 0$. We will in what follows refer to any $\bfomegab \in \Nset$ which satisfies both $\lambda_{r+1} = \ldots = \lambda_m = 0$ and $z_{r+1} = \ldots = z_m = 0$ as a rank $r$ point, even in the case that $\bfomegab \neq \Mset$. The reason for using this terminology is that it is often difficult to verify that~\eqref{eq:svdset-bound} is satisfied but sufficient to provide a parametrization around rank $r$ points, $\bfomegab \in \Nset$.

Let
$$
p_r \defeq m + \frac{r(r+1)}{2} + 1
$$
and
$$
q_r \defeq r(m+2)
$$
and note that $p = p_m$ and $q = q_m$. Further, let $\bfu_k$ denote the $k$th column of $\bfU$. It will in what follows be shown that $\bfomega$, in a neighborhood of a rank $r$ point, $\bfomegab$, can be parameterized by specifying $\lambda_k$ and $z_k$ for $k=r+1,\ldots,m$, a subset of $m-k$ parameters from $\bfu_k$ for $k=r+1,\ldots,m$, and a subset of $q_r-p_r$ parameters from
$$
\bfomega_r \defeq
(\bfu_1,\ldots,\bfu_r,\lambda_1,\ldots,\lambda_r,z_1, \ldots,z_r).
$$
It is straightforward to verify that this amounts to a total of $q-p$ parameters. The specific parameters chosen from $\bfu_k$ for $k=r+1,\ldots,m$ and from $\bfomega_r$ will remain unspecified. In line with the previous discussion these must ultimately depend on the specific $\bfomegab$ around which $\Mset$ or $\Nset$ is parameterized.

Before proving the preceding statement consider first the slightly more general system of equations given by
\begin{subequations} \label{eq:svdsetr-eq}
\begin{align}
\Trace(\bfLambda_r) + \eta &=  1 \label{eq:svdsetr-trace} \\
\diag(\bfU_r\bfLambda_r\bfU_r) + \bfgamma &= \bfU_r\bfz_r \label{eq:svdsetr-range} \\
\bfU_r\tr\bfU_r &= \bfI \label{eq:svdsetr-orth}
\end{align}
\end{subequations}
where $(\bfU_r,\bflambda_r,\bfz_r,\bfgamma,\eta) \in \reals^{m \times r} \times \reals^r \times \reals^r \times \reals^m \times \reals^1$ for some $r$, $1 \leq r \leq m$. For now, it is sufficient to view the addition of $\bfgamma$ and $\eta$ as (small) perturbations of the constraints in~\eqref{eq:svdsetr-eq}. These will later be used to develop a perturbation analysis of the solutions to~\eqref{eq:svdset-eq} around the rank $r$ points.

Let
$$
\bfomega_r \defeq (\bfU_r,\bflambda_r,\bfz_r)
$$
and define $\bfomegab_r$ analogously. Define
$$
H_r : \reals^{q_r+m+1} \mapsto \reals^{p_r}
$$
according to
$$
H_r(\bfomega_r,\bfgamma,\eta) \defeq \begin{bmatrix}
\Trace(\bfLambda_r^2) + \eta - 1 \\
\diag(\bfU_r\bfLambda_r^2\bfU_r\tr) + \bfgamma - \bfU_r\bfz_r \\
\svec(\bfU_r\tr\bfU_r-\bfI) \\
\end{bmatrix}
$$
and note that $H_r(\bfomega_r,\bfgamma,\eta) = \bfZero$ is equivalent to~\eqref{eq:svdsetr-eq}. In order to establish that the solution set of~\eqref{eq:svdsetr-eq} can (locally around a particular solution ($\bfomegab_r$,$\bfZero$,$0$)) be parameterized by $q_r-p_r+m+1$ parameters it is sufficient to establish that the Jacobian
\begin{equation} \label{eq:Dr-definition}
\bfD_r = \left. \frac{\partial H_r(\bfomegab_r)}{\partial \bfomegab_r}\right|_{\bfomega = \bfomegab} \in \reals^{p_r \times q_r}
\end{equation}
is full rank when evaluated at $\bfomegab_r$ satisfying $H_r(\bfomegab_r,\bfZero,0) = \bfZero$.

Note that, similarly to before, if~$\bfD_r$ in~\eqref{eq:Dr-definition} is full rank then this implies the existence of a Lipschitz continuous function
\begin{equation} \label{eq:psir}
\psi_r : \Dset_r \mapsto \Rset_r
\end{equation}
where $(\bfU_r,\bflambda_r,\bfz_r) = \psi_r(\bfxi_r,\bfgamma,\eta)$ for $\bfxi_r \in \reals^{q_r-p_r}$, where $\Dset_r \in \reals^{q_r-p_r+m+1}$ is an open neighborhood of $\bfZero$, and where $\Rset_r = \varphi_r(\Dset_r)$. Also, without loss of generality it can be assumed that
$$
\Dset_r = (-\kappa,\kappa)^{q_r-p_r + m + 1}.
$$

In order to establish the full rank property of $\bfD_r$ consider the matrix
$$
\bfDt_r \defeq \left. \frac{\partial H_r(\bfomegab_r)}{\partial (\bfg_1\tr,\ldots,\bfg_m\tr,\bfz_r\tr,\bflambda_r\tr)} \right|_{\bfomega = \bfomegab}
$$
where $\bfg_k$ is the $k$th \emph{row} of $\bfU_r$, i.e.~
$$
\bfU_r = \begin{bmatrix}
  \bfu_1 & \cdots & \bfu_r \\
\end{bmatrix} = \begin{bmatrix}
  \bfg_1 & \cdots & \bfg_m \\
\end{bmatrix}\tr.
$$
Note that $\bfDt_r$ is related to $\bfD_r$ by a permutation of the columns (due to a changed order of differentiation) and that $\bfDt_r$ is full rank if and only if $\bfD_r$ is full rank. Computing $\bfDt_r$ (semi) explicitly yields
$$
\bfDt_r =
\begin{bmatrix}
\bfZero & \cdots & \bfZero & \bfZero & \bflambdab_r\tr \\
2\bfgb_1\tr\bfLambdab^2_r-\bfzb_r\tr & \cdots & \bfZero\tr & \bfgb_1\tr & 2\bfLambdab_r\bfgb_1^2 \\
\vdots & \ddots & \vdots & \vdots & \vdots \\
\bfZero\tr & \cdots & 2\bfg_m\tr\bfLambdab^2_r-\bfzb_r\tr & \bfgb_m\tr & 2\bfLambdab_r\bfgb_m^2\\
\bfGb_1  & \cdots &
\bfGb_m & \bfZero & \bfZero
\end{bmatrix}
$$
where
$$
\bfGb_k \defeq \left. \frac{\partial G_r(\bfU_r)}{\partial \bfg_k} \right|_{\bfomega_r = \bfomegab_r}
\text{~for} \quad
G_r(\bfU_r) \defeq \svec(\bfU_r\tr\bfU_r-\bfI)
$$
and where $\bfgb_i^2$ denotes element wise squaring of $\bfgb_i$. Assume first that $2\bfgb_i\tr\bfLambdab^2_r-\bfzb_r\tr = \bfZero$ for some $i$, $1 \leq i \leq m$. This implies through~\eqref{eq:svdsetr-range} (and $\bfgamma = \bfZero$) that
$$
\bfgb_i\tr \bfLambdab^2_r \bfgb_i = 2 \bfgb_i\tr \bfLambdab^2_r \bfgb_i
$$
and in turn $\bfLambdab^2_r \bfgb_i = \bfZero$ as $\bfLambda^2_r \succeq \bfZero$. Further, it follows that $\bfzb_r = \bfZero$ and that $\bfLambdab_r = \bfZero$ by inserting $\bfzb_r = \bfZero$ into~\eqref{eq:svdsetr-range}. This however violates~\eqref{eq:svdsetr-trace} and contradicts that $\bfomegab_r$ is a solution to~\eqref{eq:svdsetr-eq}. Thus, it can be assumed that $2\bfgb_i\tr\bfLambdab^2_r-\bfzb_r\tr \neq \bfZero$ for all $i = 1,\ldots,m$ which implies that the first $m+1$ rows of $\bfDt_r$ are linearly independent.

Establishing that the last $r(r+1)/2$ rows of $\bfDt_r$ are linearly independent is a standard exercise in proving that the $(m,r)$-Stiefel manifold (the set of $m$ by $r$ unitary matrices) has dimension
$$
mr - \frac{r(r+1)}{2}
$$
which is a well known result~\cite{Boo:86}. We will for this reason not provide an explicit proof of this. In fact, the last $r(r+1)/2$ rows of $\bfDt_r$ are not only linearly independent but also orthogonal.

What now remains to be done, in order to show that $\bfDt_r$ is full rank, is to prove that none of the first $m+1$ rows can be written as a linear combination of the remaining $r(r+1)/2$ rows. For the first row, this is obvious due to the structure of $\bfDt_r$ together with $\bflambdab_r \neq \bfZero$. For the next $m$ rows the only potential problem would be if $\bfg_i = \bfZero$ for some $i$. However, as
$$
G_r(\bfU_r) = \svec(\bfU_r\tr\bfU_r-\bfI) = \sum_{i=1}^m \svec( \bfg_i \bfg_i\tr) - \svec(\bfI)
$$
it follows that $\bfGb_i$ is linear in $\bfgb_i$ and equal to zero whenever $\bfgb_i = \bfZero$. Together with the property that $2\bfgb_i\tr\bfLambdab^2_r-\bfzb_r\tr \neq \bfZero$ it follows that none of the first $m+1$ rows can be formed as a linear combination of the remaining $r(r+1)/2$ rows. This establishes that $\bfDt_r$, and $\bfD_r$, are full rank. Note that as
$$
\bfD = \bfD_m
$$
it also follows that the assertion of~\eqref{eq:jacobian-criterion} has been proven.

Consider again the parametrization of $\Nset$ around some rank $r$ $\bfomegab \in \Nset$ and consider the matrix
$$
\bfP = \left. \frac{\partial H(\bfomega)}{\partial (\bfomega_r,\bfu_{r+1},\ldots,\bfu_m)} \right|_{\bfomega=\bfomegab}.
$$
Note that $\bfP$ is nothing more than $\bfD$ with the columns corresponding to $\lambda_k$ and $z_k$ for $k = r+1,\ldots,m$ removed. It is straightforward to verify that $\bfP$ is structured as
\begin{equation} \label{eq:Pmatrix}
\bfP = \begin{bmatrix}
\bfD_r & \bfZero & \cdots & \bfZero \\
\times & \bfFb_{r+1}\tr & \cdots & \bfZero \\
\times & \times & \ddots & \vdots \\
\times & \times & \times & \bfFb_m\tr \\
\end{bmatrix}
\end{equation}
where
\begin{equation} \label{eq:Fmatrix}
\bfFb_{k} = \begin{bmatrix}
  \bfub_1 & \cdots & \bfub_{k-1} & 2\bfub_k \\
\end{bmatrix}
\end{equation}
and where $\bfub_i$ is the $i$th column of $\bfUb$ in $(\bfUb,\bflambdab,\bfzb) = \bfomegab$. The structure of~\eqref{eq:Fmatrix} follows by differentiating $\svec(\bfU_r\tr\bfU_r-\bfI)$ with respect to the $k$th column of $\bfU_r$ (remember that $\svec$ forms a vector of the \emph{upper} triangular part of its matrix argument). Note that $\bfF_k\tr \in \reals^{k \times m}$ is full rank for any $k$, $1 \leq k \leq m$, (as the rows are orthogonal) and that $\bfD_r \in \reals^{p_r \times q_r}$ is full rank as proven earlier. By considering the structure of $\bfP$ it follows that a linearly independent set of columns can be selected by choosing $p_r$ columns form the set of columns containing $\bfD_r$ and $k$ columns from each set containing $\bfF_{k}$ for $k=r+1,\ldots,m$. This, as elaborated on earlier, is however equivalent to the statement that the set of solutions to~\eqref{eq:svdset-eq} can locally around $\bfomegab$ be parameterized by specifying $q_r-p_r$ parameters from $\bfomega_r$, $m-k$ parameters from $\bfu_k$ along with $\lambda_k$ and $z_k$ for $k=r+1,\ldots,m$.

Now, turn attention to the original problem posed by Lemma~\ref{lm:covering}, that is, the problem of obtaining a covering of $\Aset(\bfa,\bfb)$ defined in~\eqref{eq:Asetab} and where $\bfa = (a_1,\cdots,a_m)$, $\bfb = (b_1,\cdots,b_m)$ and $0 \leq b_1 \leq \ldots \leq b_m$. Let $r$ be the maximum integer for which
$$
0 = b_1 = \ldots = b_r < b_{r+1} \leq \ldots \leq b_m.
$$
As stated earlier, if $b_1 > 0$ then $\Aset(\bfa,\bfb)$ will be empty for sufficiently small $\epsilon$. It is thus safe to assume that $b_1 = 0$ and $r \geq 1$. Further, it can without loss of generality be assumed that $\epsilon$ is arbitrary small. In particular, it can be assumed that
$$
\epsilon^{\frac{b_{r+1}}{2}} < \kappa
$$
where $\kappa$ is the constant introduced in~\eqref{eq:neighborhood}.

Consider the set
$$
\Mset(\bfb) \defeq \Mset \cap \{ (\bfU,\bflambda,\bfz) ~|~ |\lambda_i| \leq \epsilon^\frac{b_i}{2} \}.
$$
The set $\Mset(\bfb)$ is chosen such that any matrix $\bfA \in \Aset(\bfa,\bfb)$ can be expressed as $\bfA = \bfU\bfLambda$ for some $(\bfU,\bflambda,\bfz) \in \Mset(\bfb)$. Thus, the parametrization of $\Mset(\bfb)$ will also provide a parametrization of $\Aset(\bfa,\bfb)$.

Let $\{ \psi^{(l)} \}_{l=1}^L$ be a set of parameterizations (around rank $r$ points) such that
\begin{equation} \label{eq:Msetcovering}
\Mset(\bfb) \subset \bigcup_{l=1}^L \Rset^{(l)}
\end{equation}
where $\Rset^{(l)} \defeq \psi^{(l)}(\Dset)$. The assumption that $\epsilon^\frac{b_{r+1}}{2} \leq \kappa$ ensures that it is suffice to consider parameterizations around rank $r$ points, $\bfomegab \in \Nset$, in order to cover $\Mset(\bfb)$. Note also that by the assumption in~\eqref{eq:neighborhood} the coordinate neighborhoods of $\psi^{(l)}$ are all equal to $\Dset$. Further, since $\Mset(\bfb) \subset \Nset$ is compact (and since $\Rset^{(l)}$ is open) it can be assumed that $L$ is finite~\cite{Rud:76}. Define $\Dset^{(l)}(\bfb)$ according to
$$
\Dset^{(l)}(\bfb) \defeq \psi^{-1}(\Mset(\bfb) \cap \Rset^{(l)})
$$
and note that $\Dset^{(l)}(\bfb) \subset \Dset$. Finally, define
$$
\Pset^{(l)}(\bfb) \defeq \{ \bfA ~|~ \exists \bfz,~(\bfU,\bflambda,\bfz) \in \Mset(\bfb) \cap \Rset^{(l)},~\bfA = \bfU\bfLambda \}
$$
where $\bfLambda \defeq \Diag(\bflambda)$ and note that
\begin{equation} \label{eq:Psetcovering}
\Aset(\bfa,\bfb) \subset \bigcup_{l=1}^L \Pset^{(l)}(\bfb).
\end{equation}

So far, the existence of a specific parametrization, given by $\Iset$, has been proven. However, not much has been said regarding the properties of this particular parametrization. Thus, to specify the benefits of the particular parametrization chosen, let in the parameter vector $\bfxi$ the components obtained by selecting a subset of $(\bfu_1,\lambda_1,z_1,\ldots,\bfu_r,\lambda_r,z_r)$ be denoted by $\bftheta_r \in \reals^{q_r-p_r}$. Similarly, let the components obtained from $\bfu_k$, for $k=r+1,\ldots,m$ be denoted by $\bftheta_k \in \reals^{m-k}$. That is,
$$
\bfxi = (\bftheta_r,\bftheta_{r+1},\lambda_{r+1},z_{r+1},\ldots,\bftheta_{m},\lambda_{m},z_{m}).
$$
Further, introduce $\bfxih$ and $\bfxit$ and partition these analogously. Assume that $\bfxi,\bfxih \in \Dset^{(l)}(\bfb)$, let $(\bfU,\bflambda,\bfz) = \psi^{(l)}(\bfxi)$ and $(\bfUh,\bflambdah,\bfzh) = \psi^{(l)}(\bfxih)$ and let $\bfA = \bfU\bfLambda$ and $\bfAh = \bfUh\bfLambdah$ where $\bfLambdah \defeq \Diag(\bflambdah)$. Further, let $\bfAt = \bfAh - \bfA$, i.e.~$\bfAt$ is the perturbation in $\bfA$ resulting from a perturbation, $\bfxit \defeq \bfxih - \bfxi$, of $\bfxi$. The objective is now to show that if $\bfxit \in \Cset$ where
\begin{align*}
\Cset \defeq \{ \bfxit ~|~ & \|\bfthetat_r\|_\infty \leq c\epsilon^\frac{1}{2},~ \|\bfthetat_k\|_\infty \leq c\epsilon^\frac{1-b_k}{2},~ |\tilde{\lambda}_k| \leq c\epsilon^\frac{1}{2},\\
& |\tilde{z_k}| \leq c\epsilon^\frac{1}{2},~ k=r+1,\ldots,m \}
\end{align*}
and $c$ is some (yet to be defined) constant it will follow that
\begin{equation} \label{eq:Aperturbation}
\|\bfAh - \bfA \| = \| \bfAt \| \leq \epsilon^\frac{1}{2}.
\end{equation}
In the above and in the following, $\hat{\lambda}_k$, $\tilde{\lambda}_{k}$, $\hat{z}_k$ and $\tilde{z}_k$ refer to the $k$th component of $\bflambdah$, $\bflambdat$, $\bfzh$ and $\bfzt$ respectively.

Let $\bfu_k$ and $\bfuh_k$ denote the $k$th columns of $\bfU$ and $\bfUh$. Let
$$
(\bfUt,\bflambdat,\bfzt) = (\bfUh,\bflambdah,\bfzh) - (\bfU,\bflambda,\bfz)
$$
and let $\bfut_k$ denote the $k$th column of $\bfUt$. The first step is to prove that $\|\bfut_k\|_\infty \leq c K_k \epsilon^\frac{1-b_k}{2}$ for some constant $K_k$. Note that since $b_1 \leq \ldots \leq b_m$ it follows immediately from the Lipschitz continuity of $\psi$ that $\|\bfut_m\| \leq cK_m \epsilon^\frac{1-b_m}{2}$ for some constant $K_m$. This is since $\epsilon^\frac{1-b_k}{2} \leq \epsilon^\frac{1-b_m}{2}$ for $k \leq m$ implies that $\|\bfxit\|_\infty \leq c\epsilon^\frac{1-b_m}{2}$ and $K_m$ could simply be selected as the Lipschitz constant (in $\infty$-norm) of $\psi$.

For $k < m$, let $\bfU_k \in \reals^{m \times r}$ be the matrix consisting of the first $k$ columns of $\bfU$, let $\bflambda_k \in \reals^k$ the vector of the first $k$ elements of $\bflambda$ and let $\bfz_k \in \reals^k$ be the vector of the first $k$ elements of $\bfz$. Assume that $\|\bfut_i\| \leq cK_i \epsilon^\frac{1-b_i}{2}$ for some $k < i \leq m$ and note that $(\bfU_k,\bflambda_k,\bfz_k)$ must satisfy~\eqref{eq:svdsetr-eq} for
$$
\bfgamma = \sum_{i=k+1}^m \lambda_i^2 \diag(\bfu_i\bfu_i\tr) - \bfu_i z_i
$$
and
$$
\eta = \sum_{i=k+1}^m \lambda_i^2.
$$
Note also that, by the structure of $\bfP$ in~\eqref{eq:Pmatrix} it follows that
\begin{align} \label{eq:Ukbound}
&(\bfU_k,\bflambda_k,\bfz_k) = \nonumber \\ &\psi_k(\bftheta_r,\bftheta_{r+1},\lambda_{r+1},z_{r+1},\ldots,\bftheta_k,\lambda_k,z_k,\bfgamma,\eta)
\end{align}
where $\psi_k$ is the function given by the implicit function theorem in~\eqref{eq:psir}. By expanding
\begin{align*}
\bfgammah &\defeq \sum_{i=k+1}^m \hat{\lambda}_i^2 \diag(\bfuh_i\bfuh_i\tr) - \bfuh_i \hat{z}_i \\
&= \sum_{i=k+1}^m (\lambda_i+\tilde{\lambda}_i)^2 \diag((\bfu_i+\bfut_i)(\bfu_i+\bfut_i)\tr) \\
& - (\bfu_i+\bfut_i) (z_i+\tilde{z}_i)
\end{align*}
and
$$
\hat{\eta} \defeq \sum_{i=k+1}^m \lambda_i^2 = \sum_{i=k+1}^m (\lambda_i+\tilde{\lambda}_i)^2
$$
it is straightforward to show that $\bfgammat \defeq \bfgammah - \bfgamma$ and $\tilde{\eta} \defeq \hat{\eta} - \eta$ satisfies
$$
\|\bfgammat\|_\infty \leq c\tilde{K}_k \epsilon^\frac{1}{2}
\quad
\text{and}
\quad
|\eta| \leq c\tilde{K}_k \epsilon^\frac{1}{2}
$$
for some constant $\tilde{K}_k$. In essence, the potentially large perturbation (on the order or $\epsilon^\frac{1-b_i}{2}$) in $\bftheta_i$ for $i$, $k < i \leq m$ is always multiplied by factors on the order of $\epsilon^{\frac{b_i}{2}}$ which results in a perturbation, $\bfgammat$, on the order of $\epsilon^\frac{1}{2}$. Note also that it is implicitly assumed that $\epsilon$ is such that $c\tilde{K}_k \epsilon^\frac{1}{2} \leq \kappa$ or otherwise $(\bfomega_r,\bfgamma,\eta) \notin \Dset_r$. However, as $\epsilon$ can be assumed arbitrary small this is not a problem.

By the Lipschitz continuity of $\psi_k$ in~\eqref{eq:psir}, it follows that
$$
\| \bfut_k \|^2 \leq cK_k \epsilon^\frac{1-b_k}{2}
$$
for some constant $K_k$ since the argument in~\eqref{eq:Ukbound} is bounded by $$
\max(c\epsilon^\frac{1-b_k}{2},c\tilde{K}_k\epsilon^\frac{1}{2}) \leq c\tilde{K}_k \epsilon^\frac{1-b_k}{2}.
$$
By induction it follows that $\| \bfut_k \|^2 \leq cK_k \epsilon^\frac{1-b_k}{2}$ for $k=r+1,\ldots,m$ and $\|\bfut_k\| \leq cK_r \epsilon^\frac{1}{2}$ for $k=1,\ldots,r$ where $K_k$, $k = r,\ldots,m$, are constants independent of $\epsilon$ and $c$. Now, by expanding
\begin{align*}
\bfAh = & \bfUh\bfLambdah = (\bfU+\bfUt)(\bfLambda+\bfLambdat) \\
= & \bfU\bfLambda + \bfU \bfLambdat + \bfUt\bfLambda + \bfUt\bfLambdat
\end{align*}
it follows that $\bfAt \defeq \bfAh-\bfA$ satisfies $\|\bfAt\| \leq cK \epsilon^\frac{1}{2}$ for some constant, $K$. Finally, by selecting $c$ according to $c = K^{-1}$ it follows that
$$
\|\bfAt\| = \|\bfAh - \bfA \| \leq \epsilon^\frac{1}{2}.
$$

What has been shown so far is that a perturbation, $\bfxit$, around a point, $\bfxi$, in the parameter space $\Dset^{(l)}$ will, given that $\bfxit \in \Cset$, result in a perturbation of $\bfA$, $\bfAt$, which satisfies $\|\bfAt\| \leq \epsilon^\frac{1}{2}$. This implies that given a set of $\bfxi \in \Dset^{(l)}(\bfb)$, $\{\bfxi^{(l,i)} \}_{i=1}^I$, for which
$$
\Dset^{(l)}(\bfb) \subset \bigcup_{i=1}^I \Cset(\bfxi^{(l,i)})
$$
where
$$
\Cset(\bfxi) \defeq \Cset + \bfxi,
$$
we will also have a covering of $\Pset^{(l)}(\bfb)$ given by
\begin{equation}
\Pset^{(l)}(\bfb) \subset \bigcup_{i=1}^I \Aset_\epsilon(\bfA^{(l,i)})
\end{equation}
where $\bfA^{(l,i)} = \bfU^{(l,i)}\bfLambda^{(l,i)}$,
$$
(\bfU^{(l,i)},\bflambda^{(l,i)},\bfz^{(l,i)}) \defeq \psi^{(l)}(\bfxi^{(l,i)}),
$$
$\bfLambda^{(l,i)} \defeq \Diag(\bflambda^{(l,i)})$ and where $\Aset_\epsilon(\bfA)$ is defined in~\eqref{eq:balldef}. However, as $\Cset(\bfxi)$ is simply a (rectangular) box centered at $\bfxi$ and since
\begin{align}
\Dset^{(l)}(\bfb) \subset \{ \bfxi ~|~ & \|\bftheta_r\|_\infty \leq 2,~ \|\bftheta_k\|_\infty \leq 1,~|\lambda_k| \leq \epsilon^\frac{b_k}{2},\nonumber \\
& |z_k| \leq 2\epsilon^\frac{b_k}{2},~ k=r+1,\ldots,m \}
\end{align}
it follows that $\{\bfxi^{(l,i)}\}_{i=1}^I$ could be chosen such that
$$
I \dotleq \epsilon^{-\mu}
$$
where
$$
\mu = \frac{(q_r-p_r)}{2} + \sum_{k=r+1}^m \frac{(m-k)(1-b_k)^+}{2} + \frac{2(1-b_k)^+}{2}.
$$
This follows from the general statement that in order to cover a large $M$-dimensional box with side lengths $\epsilon^{\beta_i}$, $i = 1,\ldots,M$, with small boxes of side length $\epsilon^{\alpha_i}$, $i=1,\ldots,M$, one needs (in the $\doteq$ sense)
$$
\prod_{i=1}^M \epsilon^{-(\alpha_i-\beta_i)^+} = \epsilon^{-\sum_{i=1}^M {(\alpha_i-\beta_i)^+}}
$$
small boxes in total. Note also that if $\alpha_i < \beta_i$ the ``small'' boxes are actually wider than the large box in the $i$th dimension which is the reason for the $(\alpha_i-\beta_i)^+$ expression as opposed to $(\alpha_i-\beta_i)$.

By noting that
$$
q_r - p_r = (m+2)r - m - \frac{r(r+1)}{2} - 1 = \sum_{k=2}^r m - k + 2
$$
and using the assumption that $b_k = 0$ for $k=1,\ldots,r$ it follows that $\mu$ can be written as
$$
\mu = \sum_{k=2}^m \frac{(m-k+2)(1-b_k)^+}{2}.
$$
Thus, it has so far been shown that it is possible to cover $\Pset^{(l)}$ by $I \dotleq \epsilon^{-\mu}$ sets $\Aset_\epsilon(\bfA_i)$. By~\eqref{eq:Psetcovering} and since $L$ was finite this result extends to the covering of $\Aset(\bfa,\bfb)$. That is, it has been shown that there exists a covering, $\mathfrak{A}$, which satisfies
$$
\Aset(\bfa,\bfb) \subset \bigcup_{\bfA_i \in \mathfrak{A}} \Aset_\epsilon(\bfA_i)
$$
and
$$
| \mathfrak{A} | \dotleq \epsilon^{-\mu}
$$
as was asserted by Lemma~\ref{lm:covering}. \hfill $\blacksquare$

\end{appendices}

\bibliographystyle{IEEEtran}
\bibliography{refs}


\newpage \pagestyle{empty}

\newpage

\newpage

\newpage

\newpage

\newpage

\end{document}